\documentclass{article}
\usepackage{authblk}
\usepackage{graphicx}
\usepackage{amsmath}
\usepackage{amssymb}
\usepackage{color}
\usepackage[super,sort&compress,comma]{natbib}

\title{To Blink or not to Blink? A new criterion to quantify luminescence blinking.}
\author{Eduard A. Podshivaylov}
\author{Pavel A. Frantsuzov\thanks{Corresponding author. E-mail: frantsuzov@kinetics.nsc.ru}}
\affil{Voevodsky Institute of Chemical Kinetics and Combustion SB RAS, 630090, Novosibirsk, Russia}
\date{}
\begin{document}

\maketitle

\begin{abstract}
Colloidal semiconductor quantum dots are promising materials for numerous applications due to their tunable emission and high quantum yields. However, emission instability in the form of luminescence blinking remains a significant obstacle to their practical implementation. While recent advances in synthesis and surface engineering have demonstrated partial or complete blinking suppression, the quantitative assessment of the "quality" of this suppression remains challenging. Existing criteria typically rely on threshold-based classification into ON and OFF states, which becomes inherently ambiguous owing to the continuous distribution of emission intensities in single quantum dots. Here, we propose a threshold-free two-dimensional quantitative criterion based on the variance of the recombination rate and the relative quantum yield, both of which can be extracted from standard time-correlated single-photon counting measurements. We validate the criterion using simulated blinking trajectories covering all major blinking mechanisms and demonstrate that it cleanly separates them into distinct regions of the two-dimensional parameter space while providing a quantitative measure of the degree of suppression. This approach directly quantifies the temporal stability of emission without requiring arbitrary state definitions, thereby providing a robust metric for comparing different blinking suppression strategies.
\end{abstract}

\section*{Introduction}

Colloidal semiconductor quantum dots (QDs) and nanocrystals (NCs) are promising materials for a wide range of applications, including lasers\cite{zhang2021,park2021,zhukov2021}, detectors\cite{livache2019,heo2018,balitskii2024,YannHeng2025}, single-photon sources\cite{rakhlin2023,arakawa2020,senellart2017,ryu2025}, photovoltaic cells\cite{Semonin2012, kovalenko2015,Carey2015,Duan2021,Liu2022,Hao2024}, fluorescent markers\cite{Thoulouli2008,Jin2011,Ahmad2025,doNascimento2019,Huang2024}, sensors\cite{Frasco2009,Chern2019,Lesiak2019,galstyan2021}, super-radiant light sources\cite{Krieg2021,Rainò2018} and LEDs\cite{Wang2021,Moon2019,Jang2023}. This broad applicability stems from their high quantum yield, ease of synthesis, tunable emission wavelength, and narrow spectral width (see, \textit{e.g.}, reviews in Refs. \citenum{Agarwal2023,Rempel2024,Kim2025} and references therein).

However, single nanocrystals and quantum dots are known to suffer from radiation instability, including blinking\cite{Frantsuzov2008,Yang2025,Podshivaylov2023,Podshivaylov2026,Efros2016,Yuan2018,Rabouw2019}, spectral diffusion\cite{Gao2019,Podshivaylov2019,Hinterding2021,Mangnus2023,Conradt2023}, and photodegradation\cite{An2018,baitova2023,Liu2019,Wahl2022}. These effects significantly limit their practical applications\cite{gee2025}.  Consequently, a major research challenge lies in finding methods to suppress fluctuations in the photophysical response and enhance the stability of the emission yield of single QDs and NCs. A number of recent works claim suppression of luminescence blinking and/or photodegradation (see, \textit{e.g.}, reviews in Refs. \citenum{Ahmad2025} and \citenum{Yang2025}). A diverse set of strategies is being explored to suppress these effects. They range from advanced synthetic approaches that precisely control crystalline phase purity\cite{Paul2022}, core/shell structure\cite{Dennis2012,Lane2014,Tao2023,Singh2024}, and surface morphology\cite{Panda2024}, to surface engineering techniques such as passivation\cite{Hohng2004,Biswas2025,Murali2026}, annealing\cite{wang2024NR}, ligand exchange\cite{Yang2018,yang2021,Wahl2022,Takagi2023,Praneeth2024,You2025,Mi2025,Gallagher2024}, or the filling of surface trap states\cite{Chouhan2021}, and to the use of other nanoparticles in the environment\cite{Hamada2010}.

Some studies claim that blinking is completely suppressed, while others show that it persists but the total duration of the OFF states is significantly reduced. In most cases, to analyze the "quality" of blinking suppression, authors propose using threshold-based procedures, such as determining the relative ON state fraction, as well as analyzing the ON- and OFF-time distributions. This approach can suffer from several issues, the most important being the very definition of ON/OFF states and the threshold dependence of the extracted parameters.

The first issue stems from the physics of blinking itself. The cause of blinking in single nanocrystals is fluctuations in the rate of nonradiative recombination of the excited state. The blinking mechanisms differ in the pathways of non-radiative recombination.
It is widely established in the literature that the dominant mechanism of blinking is the so-called band-edge carrier (BC) or trapping mechanism (TM), proposed by Frantsuzov and Marcus in Ref. \citenum{Frantsuzov2005}. However, other pathways can also contribute, such as Auger-blinking\cite{Efros1997} (also known as A-type blinking)  and hot carrier (HC) blinking\cite{Galland2011}. A detailed overview of the underlying mechanisms can be found in Ref. \citenum{Yang2025}. This implies that the distribution of detected emission intensity is continuous, making it generally difficult to define precise ON and OFF states. Consequently, threshold selection appears to be rather arbitrary and is likely performed manually in most cases.

Second, it is well known that the procedure for constructing ON- and OFF-time distributions is highly threshold-dependent and therefore unreliable. That is, even a slight shift in the threshold position can yield different values for the characteristic times and the power-law exponent of the distribution\cite{Frantsuzov2009}. Moreover, the resulting distributions also depend on the bin size\cite{Amecke2014,Bae2016,Seth2021}.

Another option is to estimate the power spectral density (PSD) of the blinking process. As first shown by Pelton \textit{et al.} in Ref. \citenum{Pelton2004}, the characteristic PSD shape follows a power law of the form $1/f$ \cite{Frantsuzov2009,Frantsuzov2013,Podshivaylov2023,Podshivaylov2026}. It was shown by Praneeth \textit{et al.} in Ref. \citenum{Praneeth2024} that complete suppression of blinking is accompanied by a change in the shape of the PSD from a $1/f$ dependence to an almost constant value across the studied frequency range. This provides strong evidence that blinking is completely suppressed. However, such a situation is not observed when blinking persists, even with greatly reduced OFF-time durations, which is a far more commonly reported scenario.

This work is devoted to developing a quantitative criterion for evaluating the suppression of luminescence blinking, specifically in cases where OFF-state durations are reduced but not fully eliminated. To test the proposed criterion, we use simulated blinking data, based on A- and TM-type blinking mechanisms.

\section*{Theoretical background}

Let us consider an experiment on pulsed excitation of a single NC with repetition period $T_r$. The power of each pulse is $P\,[Jcm^{-2}]$.
The mean number of photons absorbed by the nanocrystal per pulse is
 $$N_a=\frac {P\sigma_a}{\hbar \omega}$$
 where $\sigma_a$ is the nanocrystal absorption cross-section and $\hbar \omega$ is the excitation photon energy.
We suppose that $N_a \ll 1$, so the probability of a double excitation is very low.

From a theoretical point of view, the PL emission can be characterized by an instant quantum yield $Y(t)$.
This quantity is defined as the probability that a nanocrystal excited at time $t$ will emit a photon.
Another characteristic of the PL is the instant relaxation rate $\Gamma(t)$ which determines
the probability distribution of the delay time $\delta t$ of the photon emitted by a nanocrystal excited at time $t$:
$$  p(\delta t|t)= \Gamma(t)\exp(-\Gamma(t) \delta t)$$
It should be noted that some nanocrystals exhibit biexponential decay caused by the presence of delayed emission.
We identify $\Gamma(t)$ with the short exponential component.

The estimators of $Y(t)$ and $\Gamma(t)$ can be extracted from the time-tagged single photon counting data: the absolute arrival times of photons $t_i$, the delay times of detected photons relative to the excitation pulse $\delta t_i$, and/or the detector number in the case of a Hanbury Brown and Twiss setup.
After preliminary processing, the arrival time data are binned to obtain the dependence of the number of photons detected per bin on time, \textit{i.e.}, the intensity trace. Thus, for each bin of size $\Delta t$ with index $j$, started at time $t_j$, a number of detected photons $n_j$ (intensity) can be assigned.
The mean number of the detected photons within the $j$-th bin is
$$N_j=   \alpha N_a \frac {\Delta t}{T_r}  \langle Y(t) \rangle_j  $$
where $\alpha$ is the efficiency of the optical system and $\langle Y(t) \rangle_j$ is the averaged quantum yield per the $j$-th bin
$$  \langle Y(t) \rangle_j = \frac 1 {\Delta t} \int\limits_{t_j}^{t_j+\Delta t} Y(t)\,dt$$
The measured intensity $n_j$ is a random number from a Poisson distribution with mean $N_j$.
Thus, the estimated mean number is $\hat N_j=n_j$ and the estimated quantum yield value is
\begin{equation}
 \hat Y_j=  \frac {n_j T_r}{\alpha N_a \Delta t}
 \label{Y_j}
\end{equation}

The estimation of $\Gamma(t)$ requires a more involved procedure. The already obtained blinking trajectory is divided into several intensity levels. For each level, all photons falling into that level are extracted, and the resulting set is then subdivided into subsets with a certain number of photons. In some cases, no subdivision by photon number is performed, and all photons belonging to the level are taken together. Subsequently, using the delay times $\delta t_i$ corresponding to the photon indices within a given level, the estimate is extracted using fitting methods such as weighted least squares or maximum likelihood. The fitting function is
\begin{equation}
    p_{K}(\delta_i) = \Gamma_K e^{-\Gamma_K \delta_i}
\end{equation}
where $\Gamma_K$ is the luminescence decay rate for the $K$-th set. As a result of the fitting, an estimate of the parameter is obtained, which we denote as $\hat{\Gamma}_{K}$. For simplicity, we will assume that each bin $j$ included in the set $K$ is assigned $\hat{\Gamma}_j = \hat{\Gamma}_K$, thus forming the set $\{\hat{\Gamma}_j\}$.

All mechanisms of luminescence blinking proposed to date, from a formal standpoint, imply that blinking leads to variations in $Y(t)$ and $\Gamma(t)$ on macroscopic time scales.
A standard way to distinguish mechanisms of blinking is the estimation of the luminescence lifetime for photons belonging to a given intensity level, i.e., the construction of fluorescence lifetime-intensity distributions (FLID).

A-type blinking implies that both $\Gamma$ and $Y$ switch between several discrete states. In the simplest case it is switching between a neutral and a negatively charged nanocrystal. The corresponding quantum yields and relaxation rates of the neutral exciton $X$ and charged exciton (trion) $X^-$ are given by the following expressions:

\begin{equation}
    \begin{aligned}
        &\Gamma_X = k_\textrm{r} + k_\textrm{nr}; \quad &&Y_X = k_\textrm{r} \Gamma_\textrm{X}^{-1}; \\
        & \Gamma_{{X}^-} = 2k_\textrm{r} + k_\textrm{nr}' + k_\textrm{A}; \quad &&Y_{{X}^-} = 2k_\textrm{r} \Gamma_{X^-}^{-1};
    \end{aligned}
    \label{eqn:Auger}
\end{equation}
where $k_\textrm{r}$ is the radiative recombination rate of the exciton, $k_\textrm{nr}$ is the nonradiative recombination rate of the exciton, $k_\textrm{nr}'$ is the nonradiative recombination rate of the trion, and $k_\textrm{A}$ is the Auger recombination rate of the trion.

The FLID for dots blinking via this mechanism appears as two or three spots, often connected by a nonlinear ridge. If $2 k_\textrm{r} \gg k_A$, one of the limiting cases may be observed - the intensity remains nearly unchanged over time, while the recombination rate varies by a factor of several times. Such behavior has been experimentally observed in Ref. \citenum{Galland2012}.

In the TM blinking mechanism it is assumed that the primary process affecting the intensity is the variation of the nonradiative trapping rate $k_\textrm{t}$, \textit{i.e.}:
\begin{equation}
\Gamma(t) = k_\textrm{r} + k_\textrm{t} (t); \quad Y(t) = \frac{k_\textrm{r}}{k_\textrm{r} + k_\textrm{t} (t)} = k_\textrm{r}/\Gamma(t)  = k_\textrm{r} \tau(t);
\label{eqn:BC}
\end{equation}
Due to the latter relation, dots blinking via this mechanism appear as a linear ridge in the FLID.

The least common mechanism is the hot carrier blinking mechanism (HC-blinking). In this mechanism, it is assumed that carriers can be captured into a metastable trap during the cooling process. If the rate of this capture varies, the emission intensity also changes. However, the luminescence decay rate remains unaffected, since trapping occurs before the carrier relaxes to the band edge:
\begin{equation}
\Gamma(t) = const;
\label{eqn:HC}
\end{equation}

Thus, the FLID for this mechanism appears as a straight line independent of the lifetime/decay rate.

\section*{Criteria}

FLID allows us to determine the blinking mechanism in a particular case. This, however, is a qualitative criterion. To characterize the suppression of blinking, quantitative criteria have to be used. Since blinking is described by variations in two quantities, $Y(t)$ and $\Gamma(t)$, two distinct criteria can be formulated.

\subsection*{Criterion I}

The first criterion characterizes fluctuations of the quantum yield.
In practice, determining the absorption cross-section of a specific nanocrystal is an extremely challenging task, as is estimating the absolute quantum yield via Eq.\eqref{Y_j}. Researchers therefore typically rely on the relative quantum yield, defined as
$$Q(t)= \frac {Y(t)}{Y_\textrm{max}}$$
where $Y_\textrm{max}$ is the maximum PL quantum yield observed during blinking. The average value of the relative quantum yield is
$$\langle Q \rangle= \frac {\langle Y \rangle}{Y_\textrm{max}}\equiv \frac {\langle N_j \rangle}{N_\textrm{max}}$$
where $N_\textrm{max}$ is the maximum mean number of detected photons per bin.
Since $\langle N_j \rangle=\langle n_j \rangle$, this quantity can be estimated as
\begin{equation}
    \langle \hat Q \rangle = \frac{\langle n_j \rangle}{\hat N_\textrm{max}}
\end{equation}
where $\hat N_\textrm{max}$ is an estimate of $N_\textrm{max}$.
The simplest possible estimate of this value is
\begin{equation}
    \hat{N}_\textrm{max}^{(0)} = {n}_\textrm{max}\equiv \max\limits_{j} n_j
\end{equation}
which yields the naive estimate of the relative quantum yield:
\begin{equation}
    \langle \hat Q \rangle^{(0)} = \frac{\langle n_j \rangle}{n_\textrm{max}}
    \label{Q0}
\end{equation}

Unfortunately, this naive estimate is biased towards smaller values, since $n_\textrm{max} \ge N_\textrm{max}$.
To address this issue, we propose a new estimate of $N_\textrm{max}$ based on extreme value theory\cite{gnedenko1992,deHaan2006,leadbetter1983}.
If the number of bins with $N_j = N_{\textrm{max}}$ is $M_\textrm{max}$, the mean value of ${n}_\textrm{max}$ is
\begin{equation}
    \mathbb{E} \left[n_\textrm{max}\right] = N_\textrm{max} + K(M_\textrm{max}) \sqrt{N_\textrm{max}}
    \label{eqn:nmax}
\end{equation}
where
\begin{equation}
    K(M_\textrm{max}) \approx \sqrt{2 \ln M_\textrm{max}}
\end{equation}
is a correction factor. The approximate expression $K(M_\textrm{max}) \approx \sqrt{2 \ln M_\textrm{max}}$ is sufficient for most practical purposes; a more precise formula is provided in Supplementary Note 1.
Assuming that $n_\textrm{max}$  is equal to $\mathbb{E} \left[n_\textrm{max}\right]$, we obtain the following self-consistent estimate for ${N}_\textrm{max}$:
\begin{equation}
    \hat{N}^{(s)}_\textrm{max} = \frac{1}{4}\left(\sqrt{K^2 + 4 n_\textrm{max}} - K\right)^2
\end{equation}
and, accordingly, the corrected estimate of the relative quantum yield:
\begin{equation}
    \boxed{\langle Q \rangle^{(s)} = \frac{\langle n_j \rangle}{\hat{N}^{(s)}_\textrm{max}}}
\end{equation}

By examining the bias of the naive estimate, the following expression can be obtained (see Supplementary Note 1 for details):
\begin{equation}
    \textrm{Bias}\left[\langle Q \rangle^{(0)}\right] = - \langle Q \rangle K \frac{\sqrt{N_\textrm{max}}}{N_\textrm{max}}
\end{equation}
It follows that the estimate is always biased downward, \textit{i.e.}:
$$ \mathbb{E}\left[\langle Q \rangle_{\textrm{est}}^{(0)}\right] \le \langle Q \rangle $$

The bias of the self-consistent estimate is far smaller. It can be shown (see Supplementary Note 1) that
\begin{equation}
    \textrm{Bias}\left[\langle Q \rangle^{(s)}\right] = \frac{\pi^2 \langle Q \rangle}{12 N_\textrm{max} \ln M_\textrm{max}} \ll \langle Q \rangle
\end{equation}
and therefore
\begin{equation}
    \left|\frac{\textrm{Bias}\left[\langle Q \rangle^{(s)}\right]}{\textrm{Bias}\left[\langle Q \rangle^{(0)}\right]}\right| = \frac{\pi^2}{12 \ln M_\textrm{max} \, K \, N_\textrm{max}^{1/2}} \ll 1
\end{equation}
Supplementary Note 1 further demonstrates that the mean-squared error of the self-consistent estimate is much smaller than that of the naive estimate whenever $M_\textrm{max} > 10$.

\subsection*{Criterion II}

The second criterion concerns fluctuations of the luminescence decay rate. Since we are not interested in the absolute value of the rate itself, the most natural measure of its spread is the variance $\sigma^2_\Gamma$. Given the estimated values $\hat{\Gamma}_j$ for each bin, the variance can be estimated in the conventional way:
\begin{equation}
    \hat{\sigma}^2_\Gamma = \frac{1}{J-1} \sum\limits_{j=1}^J \left(\hat{\Gamma}_j - \left\langle\hat{\Gamma}\right\rangle\right)^2
\end{equation}
where $J$ is the total number of bins. Because the absolute magnitude of this spread may vary from one quantum dot to another, a properly normalized criterion is required. A suitable normalization is provided by the Bhatia–Davis inequality \cite{bhatia2000}, which bounds the variance of any distribution supported on $[\Gamma_\textrm{min}, \Gamma_\textrm{max}]$:
\begin{equation}
    \sigma^2_\Gamma \le \left(\Gamma_\textrm{max} - \left\langle\Gamma\right\rangle \right)\left( \left\langle\Gamma\right\rangle - \Gamma_\textrm{min}\right)
\end{equation}
Equality holds in exactly two limiting cases: the variance is zero, which implies $\Gamma_\textrm{max} = \langle\Gamma\rangle = \Gamma_\textrm{min}$, i.e., the decay rate is constant and its distribution is a single delta function, or the distribution consists of two delta functions, corresponding to switching between two discrete decay-rate states. Accordingly, we define the following quantity:
\begin{equation}
    \boxed{\hat{\xi} = 1 - \frac{\hat{\sigma}^2_\Gamma}{\left(\hat\Gamma_\textrm{max} - \left\langle\hat\Gamma\right\rangle \right)\left( \left\langle\hat\Gamma\right\rangle - \hat\Gamma_\textrm{min}\right)}}
    \label{eqn:xi}
\end{equation}
This criterion possesses the following properties. When the decay rate is strictly constant (a single delta-function distribution), $\hat{\xi} = 1$. When the decay rate switches sharply between two discrete states (a bimodal distribution of two delta functions), the Bhatia–Davis bound is saturated and $\hat{\xi} = 0$. If the distribution is broadened but remains centered near a single value, $\hat{\xi}$ is close to 1; if it consists of two narrow but finite-width peaks, $\hat{\xi}$ is close to 0. For a complex, broadly distributed $\Gamma(t)$, the criterion takes an intermediate value between 0 and 1, reflecting the presence of significant but non-two-state fluctuations.

\subsection*{Classification of the blinking modes}

\begin{figure}[h]
\center{\includegraphics[width=.95\linewidth]{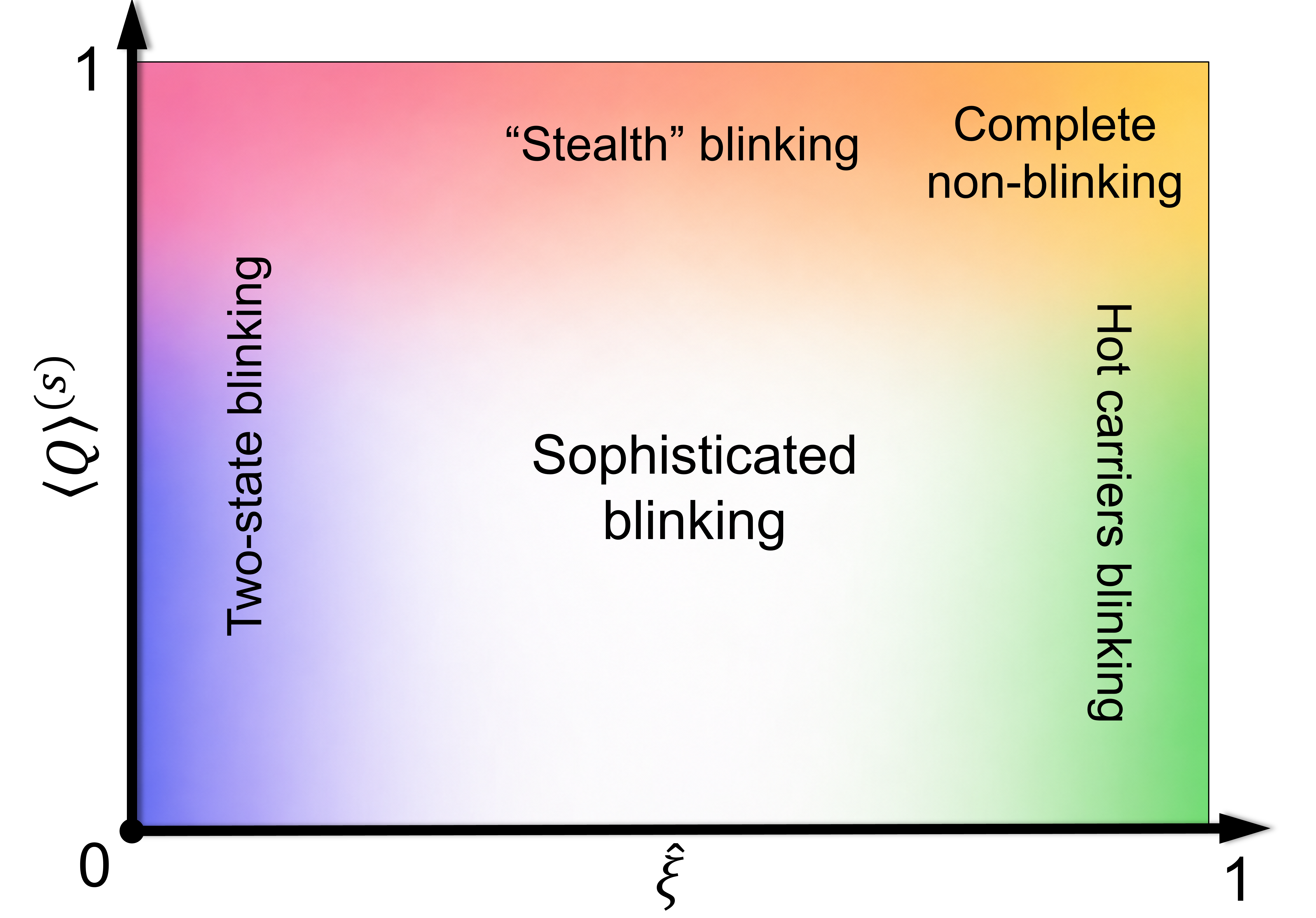}}
\caption{Two-dimensional blinking suppression criterion proposed in this work. The colored regions indicate characteristic areas corresponding to different luminescence blinking mechanisms. Blue (left): two-state blinking (A-type or two-state TM). White (center): sophisticated blinking: multilevel TM blinking and hybrid mechanisms. Green (right): hot-carrier (HC) blinking. Violet-pink (upper left):
two-state "stealth" blinking. Red (upper center): "stealth" blinking-  high relative quantum yield with persistent, non-two-state decay-rate fluctuations. Yellow-orange (upper right): complete non-blinking mode - the case of complete blinking suppression.}
\label{Fig:Classification}
\end{figure}

Summarizing the estimates proposed above, we introduce a two-dimensional criterion for blinking suppression, represented by the pair $(\hat{\xi}, \langle\hat{Q}\rangle^{(s)})$. Each quantum dot is thus mapped onto a point in the plane of decay-rate fluctuations $\hat{\xi}$ versus relative quantum yield $\langle\hat{Q}\rangle^{(s)}$. The resulting diagram, shown in \textbf{Figure \ref{Fig:Classification}}, provides a unified classification of blinking behavior and its degree of suppression.

The left side of the diagram, shaded in blue, corresponds to two-state blinking mechanisms. For both A-type blinking and two-state TM blinking, the decay rate switches between two discrete values. In the ideal limit this yields $\hat{\xi} = 0$; in practice, owing to finite sample sizes and experimental noise, the estimates form a gradient that fades rapidly towards the center of the plot. Quantum dots blinking via these mechanisms are therefore concentrated near the left edge, with their vertical position determined by their average relative quantum yield.

The central white region of the diagram corresponds to the sophisticated blinking mode: multilevel TM-blinking mechanism, as well as to hybrid scenarios involving contributions from all three mechanisms. It is in this region that the proposed criterion is most informative: rather than merely indicating the presence or absence of blinking, the coordinates $(\hat{\xi}, \langle\hat{Q}\rangle^{(s)})$ provide a quantitative measure of the degree of suppression of both intensity and decay-rate fluctuations.

The right side of the diagram, shaded in green, is associated with the hot-carrier blinking mechanism. Since HC-blinking affects only the emission intensity while the decay rate remains constant, $\hat{\xi}$ is close to 1, and the points are clustered near the right edge. Again, experimental uncertainties broaden this cluster into a gradient that decays towards the center.

The upper part of the diagram, where the relative quantum yield is high, contains three distinct regions of particular interest, corresponding to different modes of intensity stability.

In the upper-left corner, the red gradient of high quantum yield overlaps with the blue gradient of two-state decay-rate switching, producing a violet-pink shade. This corresponds to two-state "stealth" blinking  mode, reported by Galland \textit{et al.} in Ref. \citenum{Galland2012}, where nearly constant intensity was accompanied by two-level switching of the recombination rate. Here, blinking in intensity is completely suppressed ($\langle\hat{Q}\rangle^{(s)} \approx 1$), yet the decay rate continues to fluctuate between two discrete values ($\hat{\xi} \approx 0$).

The upper-central part of the diagram, shaded in red, represents a broader and, in practice, more common mode: the relative quantum yield is high, indicating strong suppression of intensity fluctuations, while the decay rate still exhibits significant variations that are not necessarily confined to two discrete states. We term this mode "stealth blinking": the nanocrystal appears stable in intensity but its excited-state lifetime continues to fluctuate.

Finally, the top right corner, shaded in yellow-orange, corresponds to the complete non-blinking mode, which is equivalent to complete blinking suppression.
 Here the green gradient of HC-blinking overlaps with the red gradient of high quantum yield, indicating that both intensity and decay-rate fluctuations are negligible ($\hat{\xi} \approx 1$, $\langle\hat{Q}\rangle^{(s)} \approx 1$).

In the following section, we perform simulations of the luminescence blinking process for the cases discussed above, accounting for realistic experimental uncertainties, and examine how reliably the proposed criterion and the associated estimates capture the blinking suppression process.

\section*{Simulation and results}

We carried out simulations for three mechanisms of blinking.

The  blinking within {\bf the  Auger mechanism}  is simulated using a model with two states: ON (neutral) and OFF (negatively charged) characterized by the pairs $\Gamma_\textrm{ON}$, $N_\textrm{ON}$ and $\Gamma_\textrm{OFF}$, $N_\textrm{OFF}$, chosen to satisfy Eq. \eqref{eqn:Auger}. The switching between states is governed by two characteristic times, $T^{+}$ (OFF $\to$ ON) and $T^{-}$ (ON $\to$ OFF), drawn from exponential distributions, \textit{i.e.}, a stochastic two-level system (TLS). The resulting telegraph process yields a sequence of switching points; between these points, the intensity and decay rate are assumed constant. The signal is then binned in time. If one or more switching events fall within a single bin, the bin intensity is taken as the time-averaged value over that bin. The corresponding decay rate for the bin is obtained by weighting the ON- and OFF-state rates by their respective intensity contributions within the bin, ensuring that the brighter state contributes proportionally more to the effective $\Gamma_j$, as would be the case in a real FLID-based analysis. Finally, Poisson noise is added to the binned intensities to mimic photon-counting statistics, producing the simulated trace $n_j$.

To account for the uncertainty inherent in estimating the decay rate from finite photon numbers, we proceed as follows. For each $\Gamma_j$, we draw a random sample $\varepsilon$ from the standard normal distribution $\mathcal{N}(0,1)$, multiply it by $\Gamma_j / \sqrt{N_\textrm{fit}}$, and apply the resulting shift to the true value. Here $N_\textrm{fit}$ is the typical number of photons used in a maximum-likelihood fit; we set $N_\textrm{fit} = 1000$ throughout. We note that the $\pm 1/\sqrt{N_\textrm{fit}}$ scaling was originally derived for the variance of lifetime estimates in Ref. \citenum{Podshivaylov2023}; the analogous form for the decay rate follows directly and is applied here. This procedure broadens the $\Gamma_j$ distribution in a manner that mimics experimental data, yielding the estimated values $\hat{\Gamma}_j$. The HC mechanism is simulated using the same TLS framework, with the simplification that the decay rate remains unchanged upon switching; only the intensity fluctuates, while the same $\Gamma$-broadening is retained. Further details of all simulations can be found in Supplementary Note 2.

\begin{figure}[h]
\centering
\includegraphics[width=0.99\linewidth]{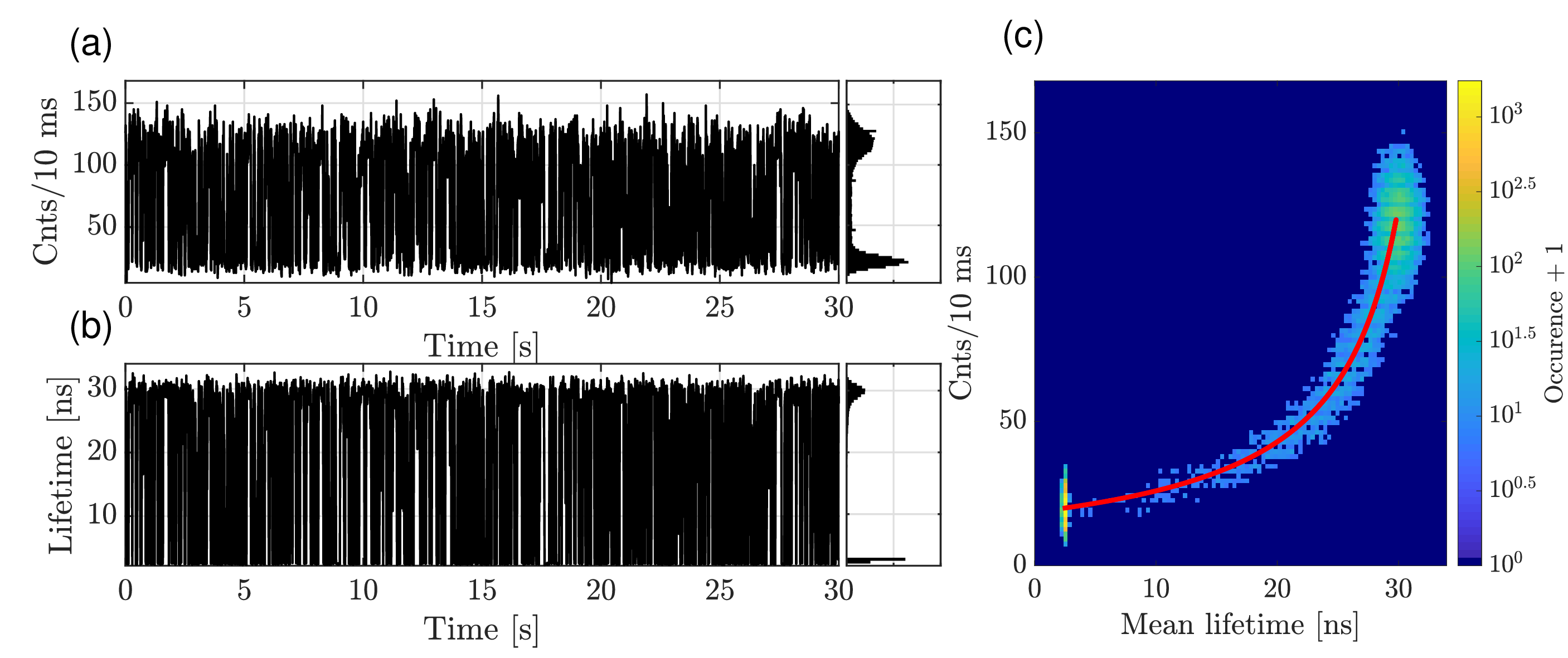}
\caption{Simulated trajectories of (a) intensity and (b) lifetime for A-type blinking with corresponding distributions. (c) FLID extracted from the
simulation. The solid red curve represents the theoretical ridge line. Note that the color scale is displayed on a logarithmic scale. Additionally, for contrast, points with fewer than 5 events have been removed.  The simulation was run for up to $10^4$ switching events with $T^{+} = T^{-} = 30$ ms and a bin size of 10 ms; only the first 30 s are shown.}
\label{Fig:Auger_trajectory}
\end{figure}

A typical simulated trace for A-type blinking is shown in \textbf{Figure \ref{Fig:Auger_trajectory}}. Each simulation was run for up to $10^4$ switching events; for clarity, only the first 30 s of the trace are displayed in the figure. Panel (a) displays the intensity trace and its distribution; panel (b) shows the corresponding lifetime trace and its distribution. The intermediate intensity and lifetime values  in these traces  are due to the contribution of the bins during which the switching occurs. This also leads to the appearance of a low ridge on the FLID graph, connecting the two peaks corresponding to the neutral and charged states of the NC in panel (c). The ridge line shown as a solid red curve in panel (c) matches the well-known behavior for A-type blinking (see, \textit{e.g.}, Ref. \citenum{Yuan2018}).

For the simulation shown here, symmetric switching times were used ($T^{+} = T^{-} = 30$ ms), with a bin size of 10 ms.

We note that a more rigorous approach would draw switching times from power-law distributions, as implemented by Palstra \textit{et al.} in Ref. \citenum{Palstra2021}. However, since the present work focuses on the distributions of intensity and decay rate across bins rather than on the temporal statistics of blinking, the simpler exponential model is sufficient for our purposes.

For the simulations for {\bf the trapping mechanism}  we used the model recently proposed in Ref. \citenum{Podshivaylov2023}.
 The model assumes that the quantum dot structure evolves over time due to a discrete set of independent stochastic two-level systems. The switching of each TLS modifies the Huang-Rhys parameter $S$ for the nonradiative transition as
\begin{equation}
    S(t) = s_0 + \sum_i \sigma_i(t) s_i,
\end{equation}
where $\sigma_i(t) \in \{0,1\}$ describes the state of the $i$-th TLS, $s_0$ is the baseline Huang-Rhys parameter with all TLSs off, and $s_i$ is the contribution of the $i$-th TLS when active. According to Marcus-Jortner theory, the nonradiative capture rate depends on $S(t)$ as
\begin{equation}
    k_\textrm{nr}(t) = k_0 S^{\alpha}(t),
\end{equation}
where $k_0$ is a pre-exponential factor and $\alpha$ is the effective number of phonons involved in the transition. The switching times of individual TLSs are assumed to be broadly distributed (log-uniform), which naturally reproduces the characteristic features of single-QD blinking. Given the time sequence $S(t)$, both the decay rate $\Gamma(t)$ and the mean bin intensity $N_j$ follow from Eq. \eqref{eqn:BC}. Poisson noise and the $\Gamma$-broadening procedure described above are then applied to obtain the simulated trajectories $n_j$ and $\hat{\Gamma}_j$. A detailed description of the simulation algorithm and the choice of parameters is provided in Supplementary Note 2.

\begin{figure}[h]
\centering
\includegraphics[width=0.99\linewidth]{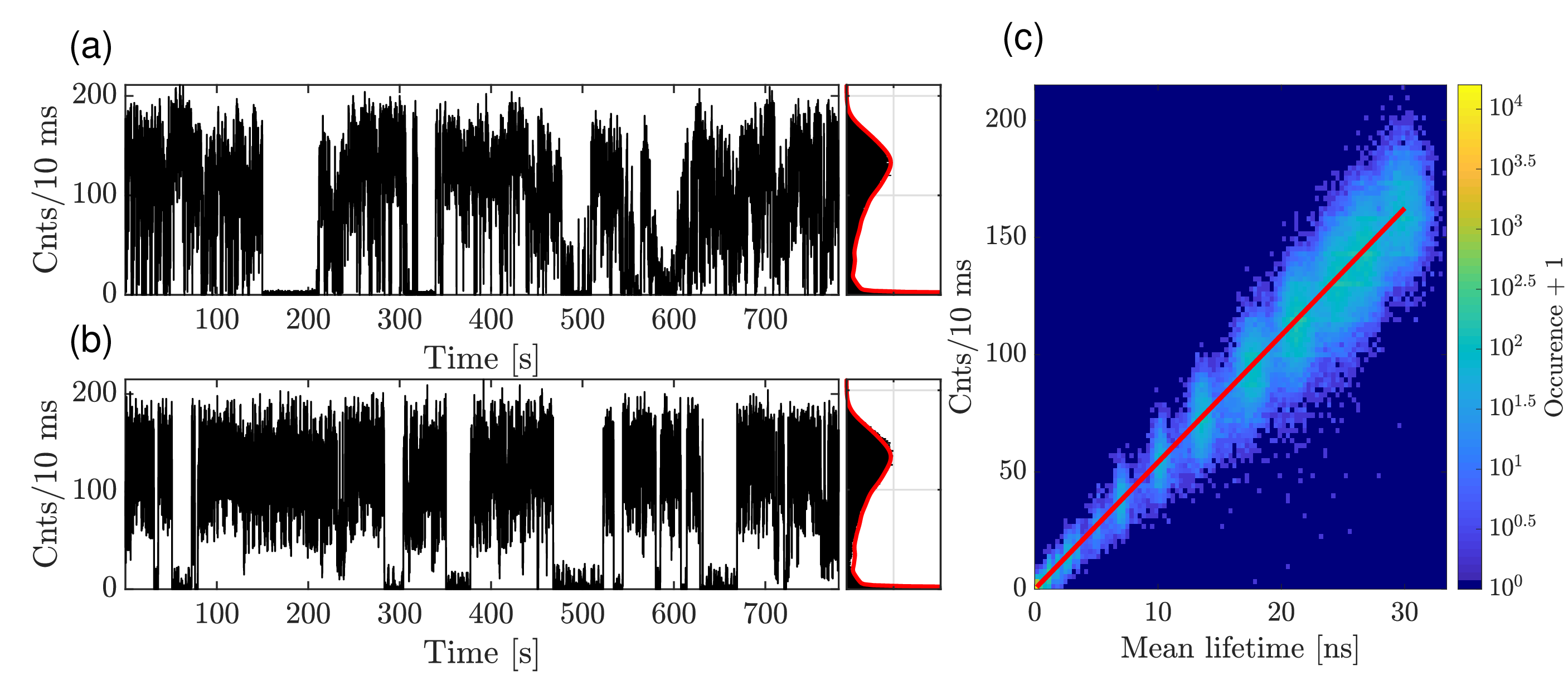}
\caption{TM blinking simulated using the model of Ref. \citenum{Podshivaylov2023}. (a) Experimental blinking trace and photon-number distribution for a single CdSeS/ZnS quantum dot (QD 3.2 in Ref. \citenum{Podshivaylov2023}); the red curve is the model fit. (b) Simulated trace and photon-number distribution generated with the parameters extracted from the fit; the red curve shows the theoretical expectation. (c) FLID extracted from the simulation, demonstrating the linear dependence characteristic of TM blinking. Simulation parameters and additional diagnostics are given in Supplementary Note 2.}
\label{Fig:BC_trajectory}
\end{figure}

To illustrate this approach, we use experimental data for a single CdSeS/ZnS core/shell quantum dot (QD 3.2 in Ref. \citenum{Podshivaylov2023}). \textbf{Figure \ref{Fig:BC_trajectory}(a)} shows the experimental blinking trace together with the photon distribution function (PDF); the red curve is the fit obtained from the model. \textbf{Figure \ref{Fig:BC_trajectory}(b)} displays a simulated trace generated using the model with the parameters extracted from that fit, along with the resulting photon distribution and the corresponding theoretical curve (red line). \textbf{Figure \ref{Fig:BC_trajectory}(c)} presents the FLID extracted from the simulation, which exhibits the linear dependence characteristic of TM blinking. The simulation preserves the known properties of all these quantities: the PDF, the FLID, the ON/OFF time distributions, the long-time autocorrelation function, and the power spectral density all exhibit the expected behavior characteristic of TM blinking. Representative examples are provided in Supplementary Note 2.

Before applying the proposed criterion to the full two-dimensional classification, we first assess the performance of the self-consistent estimator of the relative quantum yield. The two simulation approaches described above allow us to compare $\langle\hat{Q}\rangle^{(s)}$ with the naive estimator $\langle\hat{Q}\rangle^{(0)}$ across a range of blinking scenarios, using a level of accuracy that closely mimics experimental conditions.

\begin{figure}[h]
\centering
\includegraphics[width=0.99\linewidth]{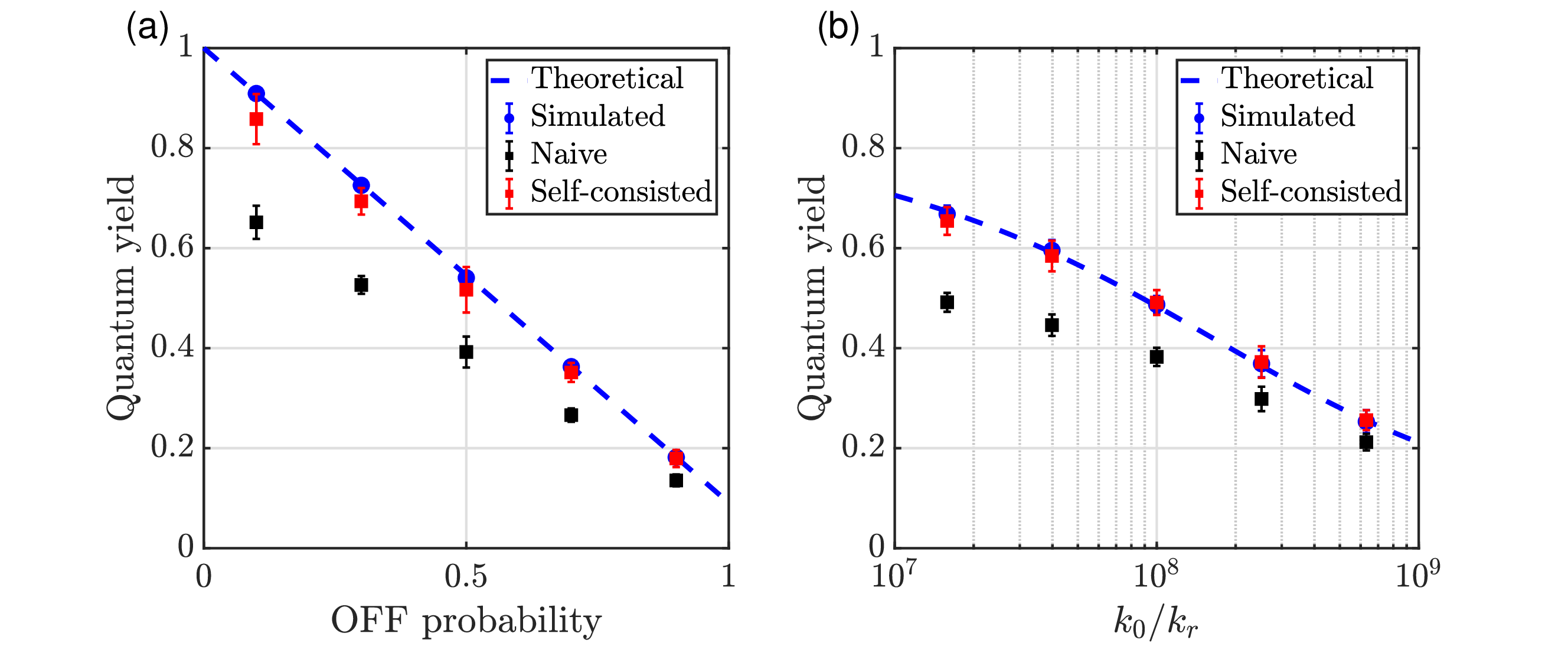}
\caption{Comparison of the naive and self-consistent estimators of the relative quantum yield. (a) A-type blinking with the OFF-state fraction $p_\textrm{OFF}$ varied from 0.1 to 0.9. Blue dashed curve: true theoretical quantum yield. Blue circles: mean of 10 simulations before Poisson noise (error bars: $\pm 1$ standard deviation). Red squares: self-consistent estimate $\langle\hat{Q}\rangle^{(s)}$. Black squares: naive estimate $\langle\hat{Q}\rangle^{(0)}$. (b) TM blinking with $k_0$ varied over several orders of magnitude (parameters from QD 3.2 of Ref. \citenum{Podshivaylov2023}). Symbols have the same meaning as in panel (a). Simulation parameters are listed in Supplementary Note 3.}\label{Fig:RelativeYieldTest}
\end{figure}

\textbf{Figure \ref{Fig:RelativeYieldTest}} presents a systematic comparison of the two estimators for both blinking mechanisms. Panel (a) shows the results for A-type blinking. The horizontal axis corresponds to the OFF-state fraction $p_\textrm{OFF} = T_\textrm{OFF}/(T_\textrm{ON}+T_\textrm{OFF})$, varied from 0.1 to 0.9. The blue dashed curve indicates the true theoretical quantum yield. Blue circles with error bars represent the mean and standard deviation of the simulated quantum yield before applying Poisson noise, averaged over 10 independent simulations at each value of $p_\textrm{OFF}$; only simulations whose quantum yield deviated from the theoretical value by less than 2\% were retained. These points lie almost exactly on the theoretical curve, confirming the fidelity of the simulation. Red squares show the self-consistent estimate $\langle\hat{Q}\rangle^{(s)}$; it agrees well with the true value, although a slight underestimation is visible at the highest quantum yields. Black squares show the naive estimate $\langle\hat{Q}\rangle^{(0)}$; it is systematically and significantly biased downward. Panel (b) presents the analogous comparison for TM blinking, using the model parameters of QD 3.2 from Ref. \citenum{Podshivaylov2023}. Here the pre-exponential factor $k_0$ is varied over several orders of magnitude to sweep the relative quantum yield. The same qualitative behavior is observed: the self-consistent estimate tracks the true value closely, while the naive estimate consistently underestimates it. All simulation parameters are provided in Supplementary Note 3.

We note that throughout this work we use the practical estimate $M_\textrm{max} \approx M_\textrm{tot}/10$, where $M_\textrm{tot}$ is the total number of time bins in the trace. As argued in Section II, the logarithmic dependence of the correction factor $K(M_\textrm{max})$ makes the criterion insensitive to the precise value of $M_\textrm{max}$, and this simple order-of-magnitude estimate is sufficient for all cases considered here. The simulations presented above confirm that this choice yields reliable results across the full range of blinking scenarios.

\begin{figure}[h]
\centering
\includegraphics[width=0.99\linewidth]{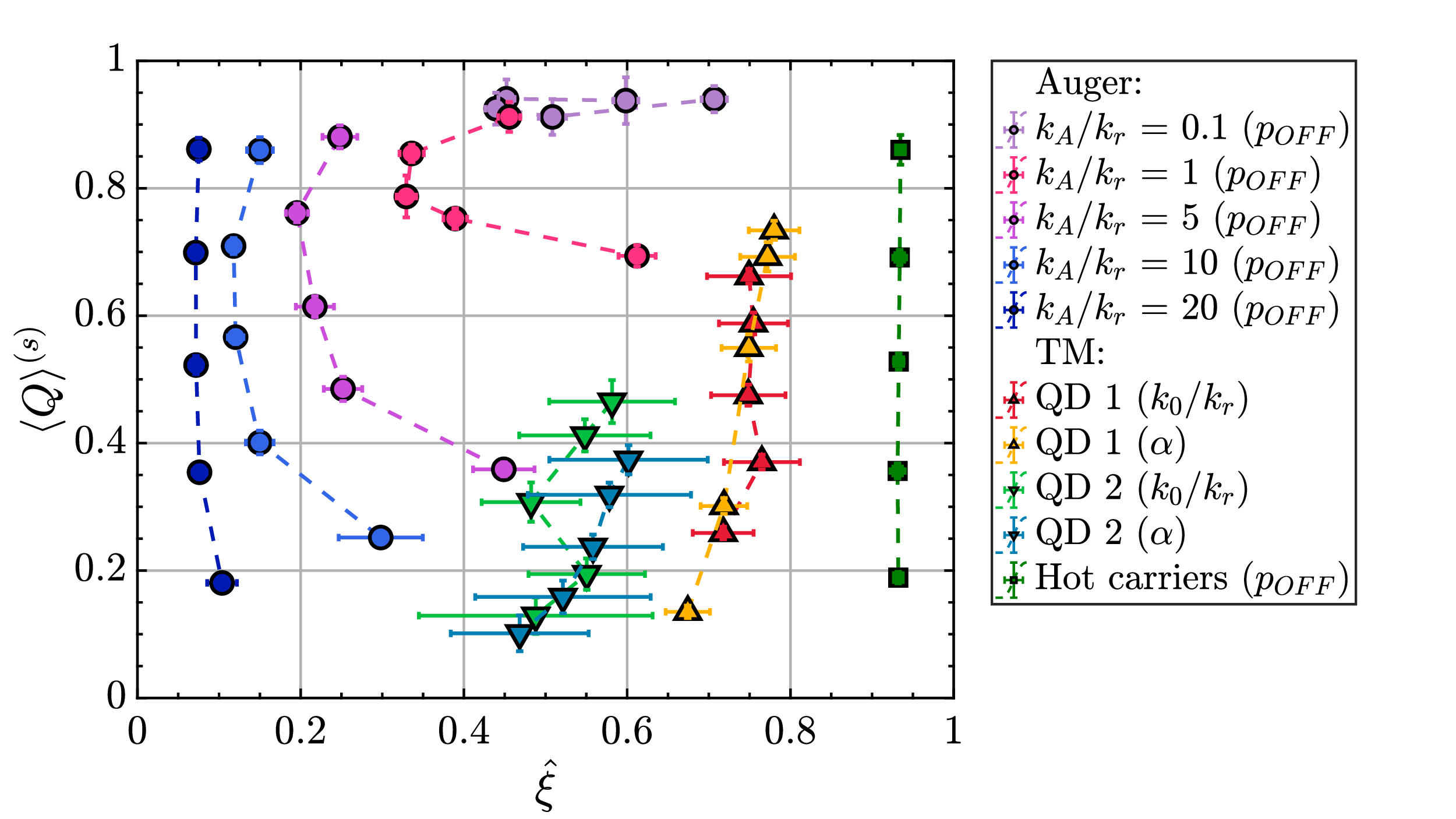}
\caption{Two-dimensional criterion $(\hat{\xi}, \langle\hat{Q}\rangle^{(s)})$ evaluated from simulations of A-type blinking (circles), TM blinking (upward triangles: QD 3.2; downward triangles: QD 4.4 from Ref. \citenum{Podshivaylov2023}), and HC blinking (green squares). For A-type blinking, five values of $k_\textrm{A}/k_\textrm{r}$ are used (0.1, 1, 5, 10, 20; see legend); for each, $p_\textrm{OFF}$ varies from 0.1 to 0.9. For TM blinking, $k_0$ is varied at fixed $\alpha=10$ (red and green) and $\alpha$ is varied at fixed $k_0$ (orange and cyan). Each point is the mean of 10 independent simulations. Simulation parameters are listed in Supplementary Note 3.}
\label{Fig:CriteriaSimulated}
\end{figure}

Having validated the self-consistent estimator, we now turn to the full two-dimensional criterion. \textbf{Figure \ref{Fig:CriteriaSimulated}} presents the results of simulations for all three blinking mechanisms, with each data point representing the mean of 10 independent simulation runs.

We begin with A-type blinking (colored circles). Five sets of simulations were performed, corresponding to five fixed values of the Auger rate $k_\textrm{A}$ (0.1, 1, 5, 10, and 20; see legend). For each $k_\textrm{A}$, the OFF-state fraction $p_\textrm{OFF}$ was varied from 0.1 to 0.9. The points lie predominantly in the left part of the diagram ($\hat{\xi} \lesssim 0.3$), consistent with the two-state nature of this mechanism. For large $k_\textrm{A}/k_\textrm{r}$ (5, 10, 20), the expected behavior is observed: $\hat{\xi}$ remains close to zero and the quantum yield decreases with increasing $p_\textrm{OFF}$. For small $k_\textrm{A}$ (0.1 and 1), however, the points do not reach the extreme upper-left corner but instead remain at intermediate $\hat{\xi}$ values. This is a consequence of the intensity-weighted averaging of the decay rate within individual time bins. When $k_\textrm{A}$ is small, the ON and OFF states have nearly equal intensities (see Eq. \ref{eqn:Auger}), so that bins containing switching events acquire intermediate $\Gamma_j$ values lying between $\Gamma_\textrm{ON}$ and $\Gamma_\textrm{OFF}$. The resulting distribution of $\hat{\Gamma}_j$ is no longer bimodal but nearly continuous, shifting $\hat{\xi}$ away from zero.

TM blinking was simulated using the model parameters of two quantum dots from Ref. \citenum{Podshivaylov2023}: QD 3.2 (upward triangles) and QD 4.4 (downward triangles). For each dot, two series of simulations were carried out. In the first series, the pre-exponential factor $k_0$ was varied over several orders of magnitude while keeping $\alpha = 10$ fixed (red and green triangles). In the second series, $k_0$ was held at the value reported in Ref. \citenum{Podshivaylov2023} while $\alpha$ was varied from 8 to 12 (orange and cyan triangles). The resulting points occupy the central region of the diagram, as expected for the multilevel TM mechanism. In contrast to A-type blinking, the scatter in $\hat{\xi}$ is substantial, reflecting the inherently broad distribution of decay rates produced by the multilevel TLS model. The relative quantum yield varies systematically with both $k_0$ and $\alpha$, correctly capturing the expected trends: $\langle\hat{Q}\rangle^{(s)}$ decreases as nonradiative recombination becomes more effective.

{ \bf HC blinking} was simulated using the same TLS framework as A-type blinking, but with the decay rate kept constant; only the intensity switches between two levels. The results are shown as green squares. The points lie on a nearly vertical line at $\hat{\xi} \approx 0.9$, with minimal scatter. The deviation from the ideal value $\hat{\xi} = 1$ is due to the $\Gamma$-broadening procedure, which introduces a small but finite variance. The horizontal spread reflects the variation of $p_\textrm{OFF}$, which controls the relative quantum yield.

Overall, the simulation results are in excellent agreement with the theoretical expectations outlined in Section II. All limiting cases identified there-two-state blinking on the left, HC blinking on the right, multilevel TM blinking in the center, and the approach to the ideal point $(1,1)$-are clearly reproduced. The criterion thus provides a quantitative and physically transparent mapping of the degree of blinking suppression across all major mechanisms. The simulation parameters are summarized in Supplementary Note 3.

\section*{Discussion}

In the preceding sections, we have proposed and tested on simulated data a new quantitative criterion for evaluating the suppression of luminescence blinking in single quantum dots. Unlike widely used threshold-based procedures, the present criterion does not rely on an arbitrary definition of ON and OFF states and, moreover, provides a physically transparent classification of the underlying blinking mechanism. The criterion is straightforward to compute. The complete procedure consists of the following steps.

\begin{enumerate}
    \item \textbf{Estimation of the relative quantum yield.} The mean intensity $\langle n_j \rangle$ and the maximum intensity $n_\textrm{max}$ are computed from the binned blinking trace. The self-consistent correction derived in Section II is applied to obtain $\langle\hat{Q}\rangle^{(s)}$. As demonstrated in the previous section, the practical choice $M_\textrm{max} \approx M_\textrm{tot}/10$ yields reliable results across all blinking modes considered.

    \item \textbf{Estimation of the decay rate and its variance.} The intensity trace is divided into a set of levels. For each level, the luminescence decay rate $\hat{\Gamma}_k$ is estimated from the delay-time data (using, \textit{e.g.}, a maximum-likelihood fit). The estimated rate is assigned to every time bin belonging to that level, producing a set $\{\hat{\Gamma}_j\}$. The mean $\langle\hat{\Gamma}\rangle$, the variance $\hat{\sigma}^2_\Gamma$, and the minimum and maximum values are computed. From these, $\hat{\xi}$ is constructed using the Bhatia-Davis normalization [Eq. \eqref{eqn:xi}]. For bins with very low intensity, several adjacent bins may need to be combined to accumulate sufficient photon statistics; the resulting $\hat{\Gamma}_k$ is then assigned to all constituent bins.

    \item \textbf{Mapping onto the $(\hat{\xi}, \langle\hat{Q}\rangle^{(s)})$ plane.} The obtained values for all QDs in the ensemble are plotted on a single diagram. If the majority of points fall into one of the characteristic regions identified in Section II (left: two-state blinking; right: HC blinking; center: multilevel TM blinking; upper regions: stealth blinking or ideal suppression), the dominant blinking mechanism can be inferred directly. If the points lie in the central region, additional FLID analysis is advisable to clarify the mechanism.

    \item \textbf{Quantification of the degree of suppression.} The positions of different ensembles are compared on the diagram. Ensembles with stronger blinking suppression shift toward the upper-right corner, approaching the ideal point $(1,1)$. This allows a fully quantitative, threshold-free comparison of the efficacy of different suppression strategies.
\end{enumerate}

A comment is in order regarding our choice to formulate the criterion in terms of the decay rate $\Gamma$ rather than the more commonly used lifetime $\tau$. In all major blinking mechanisms, it is the nonradiative recombination rate that fluctuates - whether via direct trapping (TM), Auger recombination (A-type), or hot-carrier capture (HC). Since $\Gamma = k_\textrm{r} + k_\textrm{nr}$, the differences $\Gamma_\textrm{max} - \langle\Gamma\rangle$ and $\langle\Gamma\rangle - \Gamma_\textrm{min}$ that enter the Bhatia-Davis bound depend only on the fluctuating part $k_\textrm{nr}$, with the radiative rate $k_\textrm{r}$ canceling out. The same applies to the variance $\sigma^2_\Gamma$. In contrast, the lifetime $\tau = 1/\Gamma$ mixes radiative and nonradiative contributions in a nonlinear fashion, which can introduce a bias. The criterion can, in principle, be formulated using lifetimes as well, provided this bias is properly accounted for.

The present study has been limited to the three established blinking mechanisms and their multilevel TM extension. A hybrid mechanism combining, \textit{e.g.}, Auger and TM mechanisms would require a more sophisticated simulation framework merging both approaches. The development of such a framework, and the corresponding extension of the criterion, is left for future work.

We believe that the proposed criterion offers a practical and physically grounded tool for the quantitative comparison of blinking suppression strategies, an area of intense current activity. As a next step, we intend to apply the criterion to experimental data sets, both with and without blinking suppression, to validate its performance on real nanocrystal ensembles.

\section*{Conclusion}

We have proposed a new two-dimensional criterion for evaluating the suppression of luminescence blinking in single colloidal quantum dots. The criterion combines two quantities: the self-consistent estimate of the relative quantum yield $\langle\hat{Q}\rangle^{(s)}$ and the normalized decay-rate fluctuation measure $\hat{\xi}$, both extracted directly from standard time-tagged single-photon counting data. The self-consistent estimator corrects a systematic bias inherent in the commonly used naive estimate by means of extreme-value theory, while $\hat{\xi}$ employs the Bhatia-Davis inequality to provide a bounded, normalized measure of decay-rate variations.

The performance of both estimators and of the full two-dimensional criterion has been tested on simulated blinking trajectories representing A-type, TM, and HC mechanisms, with realistic experimental noise. The self-consistent estimator consistently recovers the true relative quantum yield, whereas the naive estimator significantly underestimates it. The two-dimensional diagram $(\hat{\xi}, \langle\hat{Q}\rangle^{(s)})$ cleanly separates the three blinking mechanisms into distinct regions - left, center, and right - and identifies the modes of stealth blinking and complete blinking suppression at the top of the diagram. The criterion thus enables not only a quantitative assessment of the degree of blinking suppression but also a preliminary identification of the dominant blinking mechanism.

The proposed approach is threshold-free, computationally simple, and based solely on quantities routinely extracted from single-QD measurements. We expect it to become a useful tool for the growing community working on the stabilization of single-nanocrystal emission.

\section*{Acknowledgments}

Authors acknowledge the core funding from the Russian Federal Ministry of Science and Higher Education (FWGF-2026-0007).

\bibliography{ref}

@book{leadbetter1983,
  title     = {Extremes and Related Properties of Random Sequences and Processes},
  author    = {Leadbetter, M. R. and Lindgren, Georg and Rootz{\'e}n, Holger},
  year      = {1983},
  publisher = {Springer},
  address   = {New York, NY},
  isbn      = {978-1-4612-5451-5},
  doi       = {10.1007/978-1-4612-5449-2},
  series    = {Springer Series in Statistics},
  edition   = {1}
}

@article{bhatia2000,
  title={A better bound on the variance},
  author={Bhatia, Rajendra and Davis, Chandler},
  journal={The american mathematical monthly},
  volume={107},
  number={4},
  pages={353--357},
  year={2000},
  publisher={Taylor \& Francis}
}

@book{deHaan2006,
  title={Extreme value theory: an introduction},
  author={De Haan, Laurens and Ferreira, Ana},
  year={2006},
  publisher={Springer}
}

@incollection{gnedenko1992,
  title={On the limiting distribution of the maximum term in a random series},
  author={Gnedenko, BV},
  booktitle={Breakthroughs in Statistics: Foundations and Basic Theory},
  pages={195--225},
  year={1992},
  publisher={Springer}
}

@article{Palstra2021,
author = {Palstra, Isabelle M. and Koenderink, A. Femius},
title = {A Python Toolbox for Unbiased Statistical Analysis of Fluorescence Intermittency of Multilevel Emitters},
journal = {The Journal of Physical Chemistry C},
volume = {125},
number = {22},
pages = {12050-12060},
year = {2021},
doi = {10.1021/acs.jpcc.1c01670},
    note ={PMID: 34276862},

URL = {

        https://doi.org/10.1021/acs.jpcc.1c01670



},
eprint = {

        https://doi.org/10.1021/acs.jpcc.1c01670



}

}

@article{Galland2012,
  title={Lifetime blinking in nonblinking nanocrystal quantum dots},
  author={Galland, Christophe and Ghosh, Yagnaseni and Steinbr{\"u}ck, Andrea and Hollingsworth, Jennifer A and Htoon, Han and Klimov, Victor I},
  journal={Nature communications},
  volume={3},
  number={1},
  pages={908},
  year={2012},
  publisher={Nature Publishing Group UK London}
}

@article{Murali2026,
author = {Murali, Rahul and Panda, Mrinal Kanti and Challa, Rajendra Kumar and Acharjee, Debopam and Soma, Venugopal Rao and Ghosh, Subhadip and Raavi, Sai Santosh Kumar},
title = {Bright and Stable Single-Photon Emission in Zinc-Alloyed CsPbBr3 Nanocrystals Through Controlled Auger Recombination},
journal = {Small},
volume = {22},
number = {1},
pages = {e05011},
doi = {https://doi.org/10.1002/smll.202505011},
url = {https://onlinelibrary.wiley.com/doi/abs/10.1002/smll.202505011},
eprint = {https://onlinelibrary.wiley.com/doi/pdf/10.1002/smll.202505011},
year = {2026}
}

@article{wang2024NR,
  title={Single-particle photoluminescence connects thermal processing with heterogeneity in the trap distribution of cesium lead bromide nanocrystals},
  author={Wang, Dong and Chen, Jie and Zhang, Dongyan and Niedzwiedzki, Dariusz M and Loomis, Richard A and Sadtler, Bryce},
  journal={Nano Research},
  volume={17},
  number={12},
  pages={10363--10375},
  year={2024},
  publisher={Springer}
}

@article{Gallagher2024,
author = {Gallagher, Shaun and Kline, Jessica and Jahanbakhshi, Farzaneh and Sadighian, James C. and Lyons, Ian and Shen, Gillian and Hammel, Benjamin F. and Yazdi, Sadegh and Dukovic, Gordana and Rappe, Andrew M. and Ginger, David S.},
title = {Ligand Equilibrium Influences Photoluminescence Blinking in CsPbBr3: A Change Point Analysis of Widefield Imaging Data},
journal = {ACS Nano},
volume = {18},
number = {29},
pages = {19208-19219},
year = {2024},
doi = {10.1021/acsnano.4c04968},
    note ={PMID: 38982590},

URL = {

        https://doi.org/10.1021/acsnano.4c04968



},
eprint = {

        https://doi.org/10.1021/acsnano.4c04968



}

}

@article{Pelton2004,
    author = {Pelton, Matthew and Grier, David G. and Guyot-Sionnest, Philippe},
    title = {Characterizing quantum-dot blinking using noise power spectra},
    journal = {Applied Physics Letters},
    volume = {85},
    number = {5},
    pages = {819-821},
    year = {2004},
    month = {08},
    issn = {0003-6951},
    doi = {10.1063/1.1779356},
    url = {https://doi.org/10.1063/1.1779356},
    eprint = {https://pubs.aip.org/aip/apl/article-pdf/85/5/819/18592244/819_1_online.pdf},
}

@article{Seth2021,
author = {Seth, Sudipta and Podshivaylov, Eduard A. and Li, Jun and Gerhard, Marina and Kiligaridis, Alexander and Frantsuzov, Pavel A. and Scheblykin, Ivan G.},
title = {Presence of Maximal Characteristic Time in Photoluminescence Blinking of MAPbI3 Perovskite},
journal = {Advanced Energy Materials},
volume = {11},
number = {44},
pages = {2102449},
doi = {https://doi.org/10.1002/aenm.202102449},
url = {https://advanced.onlinelibrary.wiley.com/doi/abs/10.1002/aenm.202102449},
eprint = {https://advanced.onlinelibrary.wiley.com/doi/pdf/10.1002/aenm.202102449},
year = {2021}
}

@article{Bae2016,
author = {Bae, Youn Jue and Gibson, Natalie A. and Ding, Tina X. and Alivisatos, A. Paul and Leone, Stephen R.},
title = {Understanding the Bias Introduced in Quantum Dot Blinking Using Change Point Analysis},
journal = {The Journal of Physical Chemistry C},
volume = {120},
number = {51},
pages = {29484-29490},
year = {2016},
doi = {10.1021/acs.jpcc.6b09780},

URL = {

        https://doi.org/10.1021/acs.jpcc.6b09780



},
eprint = {

        https://doi.org/10.1021/acs.jpcc.6b09780



}

}

@article{Amecke2014,
    author = {Amecke, Nicole and Heber, André and Cichos, Frank},
    title = {Distortion of power law blinking with binning and thresholding},
    journal = {The Journal of Chemical Physics},
    volume = {140},
    number = {11},
    pages = {114306},
    year = {2014},
    month = {03},
    issn = {0021-9606},
    doi = {10.1063/1.4868252},
    url = {https://doi.org/10.1063/1.4868252},
    eprint = {https://pubs.aip.org/aip/jcp/article-pdf/doi/10.1063/1.4868252/15475516/114306_1_online.pdf},
}

@article{Frantsuzov2009,
  title = {Model of Fluorescence Intermittency of Single Colloidal Semiconductor Quantum Dots Using Multiple Recombination Centers},
  author = {Frantsuzov, Pavel A. and Volkán-Kacsó, Sándor and Jankó, Bolizsár},
  journal = {Phys. Rev. Lett.},
  volume = {103},
  issue = {20},
  pages = {207402},
  numpages = {4},
  year = {2009},
  month = {Nov},
  publisher = {American Physical Society},
  doi = {10.1103/PhysRevLett.103.207402},
  url = {https://link.aps.org/doi/10.1103/PhysRevLett.103.207402}
}

@Article{Frantsuzov2005,
  author = 	 {P. A. Frantsuzov and R. A. Marcus},
  title = 	 {Explanation of quantum dot blinking without the long-lived trap hypothesis},
  journal =  {Phys. Rev. B.},
  year = 	 {2005},
  volume = 	 {72},
  number =   {15},
  pages = 	 {155321}
}

@article{Hamada2010,
author = {Hamada, Morihiko and Nakanishi, Shunsuke and Itoh, Tamitake and Ishikawa, Mitsuru and Biju, Vasudevanpillai},
title = {Blinking Suppression in CdSe/ZnS Single Quantum Dots by TiO2 Nanoparticles},
journal = {ACS Nano},
volume = {4},
number = {8},
pages = {4445-4454},
year = {2010},
doi = {10.1021/nn100698u},
    note ={PMID: 20731430},

URL = {

        https://doi.org/10.1021/nn100698u



},
eprint = {

        https://doi.org/10.1021/nn100698u



}

}

@article{Chouhan2021,
author = {Chouhan, Lata and Ito, Syoji and Thomas, Elizabeth Mariam and Takano, Yuta and Ghimire, Sushant and Miyasaka, Hiroshi and Biju, Vasudevanpillai},
title = {Real-Time Blinking Suppression of Perovskite Quantum Dots by Halide Vacancy Filling},
journal = {ACS Nano},
volume = {15},
number = {2},
pages = {2831-2838},
year = {2021},
doi = {10.1021/acsnano.0c08802},
    note ={PMID: 33417451},

URL = {

        https://doi.org/10.1021/acsnano.0c08802



},
eprint = {

        https://doi.org/10.1021/acsnano.0c08802



}

}

@article{Yang2018,
author = {Changgang Yang and Guofeng Zhang and Liheng Feng and Bin Li and Zhijie Li and Ruiyun Chen and Chengbing Qin and Yan Gao and Liantuan Xiao and Suotang Jia},
journal = {Opt. Express},
number = {9},
pages = {11889--11902},
publisher = {Optica Publishing Group},
title = {Suppressing the photobleaching and photoluminescence intermittency of single near-infrared CdSeTe/ZnS quantum dots with p-phenylenediamine},
volume = {26},
month = {Apr},
year = {2018},
url = {https://opg.optica.org/oe/abstract.cfm?URI=oe-26-9-11889},
doi = {10.1364/OE.26.011889},
}

@article{Praneeth2024,
author = {Praneeth, N. V. S. and Akhil, Syed and Mukherjee, Amitrajit and Seth, Sudipta and Khatua, Saumyakanti and Mishra, Nimai},
title = {Amine-Free Multi-Faceted CsPbBr3 Nanocrystals for Complete Suppression of Long-Lived Dark States},
journal = {Advanced Optical Materials},
volume = {12},
number = {16},
pages = {2303222},
doi = {https://doi.org/10.1002/adom.202303222},
url = {https://advanced.onlinelibrary.wiley.com/doi/abs/10.1002/adom.202303222},
eprint = {https://advanced.onlinelibrary.wiley.com/doi/pdf/10.1002/adom.202303222},
year = {2024}
}

@article{Mi2025,
  title={Towards non-blinking and photostable perovskite quantum dots},
  author={Mi, Chenjia and Gee, Gavin C and Lander, Chance W and Shin, Donghoon and Atteberry, Matthew L and Akhmedov, Novruz G and Hidayatova, Lamia and DiCenso, Jesse D and Yip, Wai Tak and Chen, Bin and others},
  journal={Nature communications},
  volume={16},
  number={1},
  pages={204},
  year={2025},
  publisher={Nature Publishing Group UK London}
}

@article{You2025,
author = {You, Huangpeng and Liu, Zhe and Zhu, Kaijie and Wang, Qingyu and Liu, Zifeng and Li, Peixian and Hu, Xingyu and Duan, Junli and Li, Yang and Dai, Ning and Hou, Xiaoqi},
title = {Unraveling the Role of Ligands Adsorption/Desorption on Photoluminescence Blinking in Single Water-Soluble InP-Based Quantum Dots},
journal = {Nano Letters},
volume = {25},
number = {33},
pages = {12446-12454},
year = {2025},
doi = {10.1021/acs.nanolett.5c01971},
    note ={PMID: 40772843},

URL = {

        https://doi.org/10.1021/acs.nanolett.5c01971



},
eprint = {

        https://doi.org/10.1021/acs.nanolett.5c01971



}

}

@Article{Takagi2023,
author ="Takagi, Toranosuke and Omagari, Shun and Vacha, Martin",
title  ="Suppression of blinking in single CsPbBr3 perovskite nanocrystals through surface ligand exchange",
journal  ="Phys. Chem. Chem. Phys.",
year  ="2023",
volume  ="25",
issue  ="28",
pages  ="19004-19012",
publisher  ="The Royal Society of Chemistry",
doi  ="10.1039/D3CP01844J",
url  ="http://dx.doi.org/10.1039/D3CP01844J"}

@article{Biswas2025,
author = {Biswas, Subarna and Panda, Mrinal Kanti and Chatterjee, Shovon and Satra, Jit and Sharma, Shilendra Kumar and Rath, Jyotisman and Dutta, Abhijit and Acharjee, Debopam and Chakraborty, Sudip and Ghosh, Subhadip and Mishra, Nimai},
title = {Synergistic Mitigation of Phase Segregation and Blinking Suppression Along with Enhanced Electrocatalytic Activity in CsPbBrI2 Perovskite Nanocrystals via Ascorbic Acid Surface Treatment},
journal = {Advanced Functional Materials},
volume = {35},
number = {43},
pages = {2505506},
doi = {https://doi.org/10.1002/adfm.202505506},
url = {https://advanced.onlinelibrary.wiley.com/doi/abs/10.1002/adfm.202505506},
eprint = {https://advanced.onlinelibrary.wiley.com/doi/pdf/10.1002/adfm.202505506},
year = {2025}
}

@article{Hohng2004,
author = {Hohng, Sungchul and Ha, Taekjip},
title = {Near-Complete Suppression of Quantum Dot Blinking in Ambient Conditions},
journal = {Journal of the American Chemical Society},
volume = {126},
number = {5},
pages = {1324-1325},
year = {2004},
doi = {10.1021/ja039686w},
    note ={PMID: 14759174},

URL = {

        https://doi.org/10.1021/ja039686w



},
eprint = {

        https://doi.org/10.1021/ja039686w



}

}

@article{Panda2024,
author = {Panda, Mrinal Kanti and Acharjee, Debopam and Mahato, Asit Baran and Ghosh, Subhadip},
title = {Facet Dependent Photoluminescence Blinking from Perovskite Nanocrystals},
journal = {Small},
volume = {20},
number = {33},
pages = {2311559},
doi = {https://doi.org/10.1002/smll.202311559},
url = {https://onlinelibrary.wiley.com/doi/abs/10.1002/smll.202311559},
eprint = {https://onlinelibrary.wiley.com/doi/pdf/10.1002/smll.202311559},
year = {2024}
}

@article{Lane2014,
author = {Lane, Lucas A. and Smith, Andrew M. and Lian, Tianquan and Nie, Shuming},
title = {Compact and Blinking-Suppressed Quantum Dots for Single-Particle Tracking in Live Cells},
journal = {The Journal of Physical Chemistry B},
volume = {118},
number = {49},
pages = {14140-14147},
year = {2014},
doi = {10.1021/jp5064325},
    note ={PMID: 25157589},

URL = {

        https://doi.org/10.1021/jp5064325



},
eprint = {

        https://doi.org/10.1021/jp5064325



}

}

@article{Dennis2012,
author = {Dennis, Allison M. and Mangum, Benjamin D. and Piryatinski, Andrei and Park, Young-Shin and Hannah, Daniel C. and Casson, Joanna L. and Williams, Darrick J. and Schaller, Richard D. and Htoon, Han and Hollingsworth, Jennifer A.},
title = {Suppressed Blinking and Auger Recombination in Near-Infrared Type-II InP/CdS Nanocrystal Quantum Dots},
journal = {Nano Letters},
volume = {12},
number = {11},
pages = {5545-5551},
year = {2012},
doi = {10.1021/nl302453x},
    note ={PMID: 23030497},

URL = {

        https://doi.org/10.1021/nl302453x



},
eprint = {

        https://doi.org/10.1021/nl302453x



}

}

@article{Singh2024,
author = {Singh, Rahul and Praneeth, NVS and Biswas, Subarna and Palabathuni, Manoj and Muralidharan, Anandu and Mishra, Nimai and Khatua, Saumyakanti},
title = {Understanding the Size-Dependent Photostability and Photoluminescence Intermittency of Blue-Emitting Core/Graded Alloy/Shell вЂњgiantвЂќ-Quantum Dots},
journal = {Advanced Optical Materials},
volume = {12},
number = {30},
pages = {2401132},
doi = {https://doi.org/10.1002/adom.202401132},
url = {https://advanced.onlinelibrary.wiley.com/doi/abs/10.1002/adom.202401132},
eprint = {https://advanced.onlinelibrary.wiley.com/doi/pdf/10.1002/adom.202401132},
year = {2024}
}

@article{Tao2023,
author = {Tao, Chen-Lei and Ma, Jinling and Wei, Changting and Xu, Dan and Xie, Zhuoyi and Jiang, Zhengfei and Ge, Feiyue and Zhang, Han and Xie, Mingcai and Ye, Zhiliang and Cheng, Fang and Xu, Bo and Tian, Yuxi and Wu, Xue-Jun},
title = {Scalable Synthesis of High-Quality Core/Shell Quantum Dots With Suppressed Blinking},
journal = {Advanced Optical Materials},
volume = {11},
number = {17},
pages = {2300533},
doi = {https://doi.org/10.1002/adom.202300533},
url = {https://advanced.onlinelibrary.wiley.com/doi/abs/10.1002/adom.202300533},
eprint = {https://advanced.onlinelibrary.wiley.com/doi/pdf/10.1002/adom.202300533},
year = {2023}
}

@article{Paul2022,
author = {Paul, Sumanta and Samanta, Anunay},
title = {Phase-Stable and Highly Luminescent CsPbI3 Perovskite Nanocrystals with Suppressed Photoluminescence Blinking},
journal = {The Journal of Physical Chemistry Letters},
volume = {13},
number = {25},
pages = {5742-5750},
year = {2022},
doi = {10.1021/acs.jpclett.2c01463},
    note ={PMID: 35713649},

URL = {

        https://doi.org/10.1021/acs.jpclett.2c01463



},
eprint = {

        https://doi.org/10.1021/acs.jpclett.2c01463



}

}

@article{gee2025,
  title={Blinking compromises the single-photon purity of individual CsPbBr3 perovskite nanocrystals: GC Gee et al.},
  author={Gee, Gavin C and Mi, Chenjia and LaSala, Michael P and Yip, Wai Tak and Dong, Yitong},
  journal={MRS communications},
  pages={1--7},
  year={2025},
  publisher={Springer}
}

@Article{Wahl2022,
author ="Wahl, Jan and Haizmann, Philipp and Kirsch, Christopher and Frecot, Rene and Mukharamova, Nastasia and Assalauova, Dameli and Kim, Young Yong and Zaluzhnyy, Ivan and Chassé, Thomas and Vartanyants, Ivan A. and Peisert, Heiko and Scheele, Marcus",
title  ="Mitigating the photodegradation of all-inorganic mixed-halide perovskite nanocrystals by ligand exchange",
journal  ="Phys. Chem. Chem. Phys.",
year  ="2022",
volume  ="24",
issue  ="18",
pages  ="10944-10951",
publisher  ="The Royal Society of Chemistry",
doi  ="10.1039/D2CP00546H",
url  ="http://dx.doi.org/10.1039/D2CP00546H"}

@article{Liu2019,
author = {Liu, Lige and Deng, Luogen and Huang, Sheng and Zhang, Pei and Linnros, Jan and Zhong, Haizheng and Sychugov, Ilya},
title = {Photodegradation of Organometal Hybrid Perovskite Nanocrystals: Clarifying the Role of Oxygen by Single-Dot Photoluminescence},
journal = {The Journal of Physical Chemistry Letters},
volume = {10},
number = {4},
pages = {864-869},
year = {2019},
doi = {10.1021/acs.jpclett.9b00143},

URL = {

        https://doi.org/10.1021/acs.jpclett.9b00143



},
eprint = {

        https://doi.org/10.1021/acs.jpclett.9b00143



}

}

@article{An2018,
author = {An, Rui and Zhang, Fengying and Zou, Xianshao and Tang, Yingying and Liang, Mingli and Oshchapovskyy, Ihor and Liu, Yuchen and Honarfar, Alireza and Zhong, Yunqian and Li, Chuanshuai and Geng, Huifang and Chen, Junsheng and Canton, Sophie E. and Pullerits, Tõnu and Zheng, Kaibo},
title = {Photostability and Photodegradation Processes in Colloidal CsPbI3 Perovskite Quantum Dots},
journal = {ACS Applied Materials \& Interfaces},
volume = {10},
number = {45},
pages = {39222-39227},
year = {2018},
doi = {10.1021/acsami.8b14480},
    note ={PMID: 30350934},

URL = {

        https://doi.org/10.1021/acsami.8b14480



},
eprint = {

        https://doi.org/10.1021/acsami.8b14480



}

}

@article{baitova2023,
  title={Evolution of the luminescence properties of single CsPbBr3 perovskite nanocrystals during photodegradation},
  author={Baitova, Valeriya Aleksandrovna and Knyazeva, Mariya Andreevna and Mukanov, IA and Tarasevich, Alexandr Olegovich and Naumov, Andrei Vitalevich and Son, Aleksandra Grigor'evna and Kozyukhin, Sergei Aleksandrovich and Eremchev, I Yu},
  journal={JETP Letters},
  volume={118},
  number={8},
  pages={560--567},
  year={2023},
  publisher={Springer}
}

@article{Gao2019,
    author = {Gao, Kang and Springbett, Helen and Zhu, Tongtong and Oliver, Rachel A. and Arakawa, Yasuhiko and Holmes, Mark J.},
    title = {Spectral diffusion time scales in InGaN/GaN quantum dots},
    journal = {Applied Physics Letters},
    volume = {114},
    number = {11},
    pages = {112109},
    year = {2019},
    month = {03},
    issn = {0003-6951},
    doi = {10.1063/1.5088205},
    url = {https://doi.org/10.1063/1.5088205},
    eprint = {https://pubs.aip.org/aip/apl/article-pdf/doi/10.1063/1.5088205/14523756/112109_1_online.pdf},
}

@article{Hinterding2021,
author = {Hinterding, Stijn O. M. and Mangnus, Mark J. J. and Prins, P. Tim and Jöbsis, Huygen J. and Busatto, Serena and Vanmaekelbergh, Daniël and de Mello Donega, Celso and Rabouw, Freddy T.},
title = {Unusual Spectral Diffusion of Single CuInS2 Quantum Dots Sheds Light on the Mechanism of Radiative Decay},
journal = {Nano Letters},
volume = {21},
number = {1},
pages = {658-665},
year = {2021},
doi = {10.1021/acs.nanolett.0c04239},
    note ={PMID: 33395305},

URL = {

        https://doi.org/10.1021/acs.nanolett.0c04239



},
eprint = {

        https://doi.org/10.1021/acs.nanolett.0c04239



}

}

@article{Conradt2023,
author = {Conradt, Frieder and Bezold, Vincent and Wiechert, Volker and Huber, Steffen and Mecking, Stefan and Leitenstorfer, Alfred and Tenne, Ron},
title = {Electric-Field Fluctuations as the Cause of Spectral Instabilities in Colloidal Quantum Dots},
journal = {Nano Letters},
volume = {23},
number = {21},
pages = {9753-9759},
year = {2023},
doi = {10.1021/acs.nanolett.3c02318},
    note ={PMID: 37871158},

URL = {

        https://doi.org/10.1021/acs.nanolett.3c02318



},
eprint = {

        https://doi.org/10.1021/acs.nanolett.3c02318



}

}

@article{Mangnus2023,
author = {Mangnus, Mark J. J. and de Wit, Jur W. and Vonk, Sander J. W. and Geuchies, Jaco J. and Albrecht, Wiebke and Bals, Sara and Houtepen, Arjan J. and Rabouw, Freddy T.},
title = {High-Throughput Characterization of Single-Quantum-Dot Emission Spectra and Spectral Diffusion by Multiparticle Spectroscopy},
journal = {ACS Photonics},
volume = {10},
number = {8},
pages = {2688-2698},
year = {2023},
doi = {10.1021/acsphotonics.3c00420},

URL = {

        https://doi.org/10.1021/acsphotonics.3c00420



},
eprint = {

        https://doi.org/10.1021/acsphotonics.3c00420



}

}

@article{Rabouw2019,
author = {Rabouw, Freddy
T. and Antolinez, Felipe V. and Brechb{\"u}hler, Raphael and Norris, David J.},
title = {Microsecond Blinking Events in the Fluorescence of Colloidal Quantum Dots Revealed by Correlation Analysis on Preselected Photons},
journal = {The Journal of Physical Chemistry Letters},
volume = {10},
number = {13},
pages = {3732-3738},
year = {2019},
doi = {10.1021/acs.jpclett.9b01348},
    note ={PMID: 31204809},

URL = {

        https://doi.org/10.1021/acs.jpclett.9b01348



},
eprint = {

        https://doi.org/10.1021/acs.jpclett.9b01348



}

}

@article{Yang2025,
author = {Yang, Changgang and Zhang, Guofeng and Li, Jialu and Chen, Ruiyun and Qin, Chengbing and Hu, Jianyong and Yang, Zhichun and Xiao, Liantuan and Jia, Suotang},
title = {Mechanisms and Suppression of Quantum Dot Blinking},
journal = {Laser \& Photonics Reviews},
volume = {19},
number = {9},
pages = {2402269},
doi = {https://doi.org/10.1002/lpor.202402269},
url = {https://onlinelibrary.wiley.com/doi/abs/10.1002/lpor.202402269},
eprint = {https://onlinelibrary.wiley.com/doi/pdf/10.1002/lpor.202402269},
year = {2025}
}

@article{Yuan2018,
author = {Yuan, Gangcheng and Gómez, Daniel E. and Kirkwood, Nicholas and Boldt, Klaus and Mulvaney, Paul},
title = {Two Mechanisms Determine Quantum Dot Blinking},
journal = {ACS Nano},
volume = {12},
number = {4},
pages = {3397-3405},
year = {2018},
doi = {10.1021/acsnano.7b09052},
    note ={PMID: 29579376},

URL = {

        https://doi.org/10.1021/acsnano.7b09052



},
eprint = {

        https://doi.org/10.1021/acsnano.7b09052



}

}

@article{Galland2011,
  author =      {C. Galland and  Y. Ghosh and A. Steinbr\"uck and  M. Sykora and J. A. Hollingsworth
 and V. I. Klimov and H. Htoon},
  title =       {Two types of luminescence blinking revealed by spectroelectrochemistry of single quantum dots},
  journal =     {Nature},
  volume =      {479},
  number =      {},
  pages =       {203--208},
  year =        {2011}
}

@article{Efros2016,
  title={Origin and control of blinking in quantum dots},
  author={Efros, Alexander L and Nesbitt, David J},
  journal={Nature nanotechnology},
  volume={11},
  number={8},
  pages={661--671},
  year={2016},
  publisher={Nature Publishing Group UK London}
}

@article{Efros1997,
  author =    {Al. L. Efros and M. Rosen},
  title =    {Random telegraph signal in the photoluminescence intensity of a single quantum dot.},
  journal =  {Phys. Rev. Lett.},
  year =      {1997},
  OPTkey =   {},
  volume =   {78},
  number =   {6},
  pages =    {1110--1113}
}

@article{Frantsuzov2013,
author = {Frantsuzov, Pavel A. and Volkán-Kacsó, Sándor and Jankó, Boldizsár},
title = {Universality of the Fluorescence Intermittency in Nanoscale Systems: Experiment and Theory},
journal = {Nano Letters},
volume = {13},
number = {2},
pages = {402-408},
year = {2013},
doi = {10.1021/nl3035674},
    note ={PMID: 23272638},

URL = {

        https://doi.org/10.1021/nl3035674



},
eprint = {

        https://doi.org/10.1021/nl3035674
}
}

@article{Podshivaylov2019,
    author = {Podshivaylov, Eduard A. and Kniazeva, Maria A. and Gorshelev, Aleksei A. and Eremchev, Ivan Yu. and Naumov, Andrei V. and Frantsuzov, Pavel A.},
    title = {Contribution of electron-phonon coupling to the luminescence spectra of single colloidal quantum dots},
    journal = {The Journal of Chemical Physics},
    volume = {151},
    number = {17},
    pages = {174710},
    year = {2019},
    month = {11},
    issn = {0021-9606},
    doi = {10.1063/1.5124913},
    url = {https://doi.org/10.1063/1.5124913},
    eprint = {https://pubs.aip.org/aip/jcp/article-pdf/doi/10.1063/1.5124913/13809129/174710_1_online.pdf},
}

@Article{Podshivaylov2023,
author ="Podshivaylov, Eduard A. and Kniazeva, Maria A. and Tarasevich, Alexander O. and Eremchev, Ivan Yu. and Naumov, Andrei V. and Frantsuzov, Pavel A.",
title  ="A quantitative model of multi-scale single quantum dot blinking",
journal  ="J. Mater. Chem. C",
year  ="2023",
volume  ="11",
issue  ="25",
pages  ="8570-8576",
publisher  ="The Royal Society of Chemistry",
doi  ="10.1039/D3TC00638G",
url  ="http://dx.doi.org/10.1039/D3TC00638G"}

@Article{Podshivaylov2026,
author ="Podshivaylov, Eduard A. and Shekhin, Alexandr M. and Kniazeva, Maria A. and Tarasevich, Alexander O. and Sapozhnikova, Elizaveta V. and Pushkarev, Anatoly P. and Eremchev, Ivan Yu. and Naumov, Andrei V. and Frantsuzov, Pavel A.",
title  ="Model of luminescence and delayed luminescence correlated blinking in single CsPbBr3 nanocrystals",
journal  ="J. Mater. Chem. C",
year  ="2026",
pages  ="-",
publisher  ="The Royal Society of Chemistry",
doi  ="10.1039/D5TC02497H",
url  ="http://dx.doi.org/10.1039/D5TC02497H"}

@article{Agarwal2023,
doi = {10.1088/2053-1591/acda17},
url = {https://doi.org/10.1088/2053-1591/acda17},
year = {2023},
month = {jun},
publisher = {IOP Publishing},
volume = {10},
number = {6},
pages = {062001},
author = {Agarwal, Kushagra and Rai, Himanshu and Mondal, Sandip},
title = {Quantum dots: an overview of synthesis, properties, and applications},
journal = {Materials Research Express}
}

@Article{Kim2025,
author ="Kim, Jae Woo and Kim, Jigeon and Woo, Ju Young and Kim, Younghoon",
title  ="A comprehensive review of core/shell nanostructures of lead-halide perovskite quantum dots for improved optoelectronic performance and stability",
journal  ="J. Mater. Chem. A",
year  ="2025",
volume  ="13",
issue  ="36",
pages  ="29706-29735",
publisher  ="The Royal Society of Chemistry",
doi  ="10.1039/D5TA03459K",
url  ="http://dx.doi.org/10.1039/D5TA03459K"}

@article{Rempel2024,
author = {Andrey A. Rempel and Oleg V. Ovchinnikov and Ilya A. Weinstein and Svetlana V. Rempel and Yulia V. Kuznetsova and Andrei V. Naumov and Mikhail S. Smirnov and Ivan Yu. Eremchev and Alexander S. Vokhmintsev and Sergey S. Savchenko},
title = {Quantum dots: modern methods of synthesis and optical properties},
journal = {Russian Chemical Reviews},
year = {2024},
volume = {93},
pages = {RCR5114},
publisher = {ANO Editorial Board of the journal Uspekhi Khimii},
month = {May},
url = {https://rcr.colab.ws/publications/10.59761/RCR5114},
number = {4},
doi = {10.59761/RCR5114}
}

@Article{Rainò2018,
author={Rainò, Gabriele
and Becker, Michael A.
and Bodnarchuk, Maryna I.
and Mahrt, Rainer F.
and Kovalenko, Maksym V.
and Stöferle, Thilo},
title={Superfluorescence from lead halide perovskite quantum dot superlattices},
journal={Nature},
year={2018},
month={Nov},
day={01},
volume={563},
number={7733},
pages={671-675},
issn={1476-4687},
doi={10.1038/s41586-018-0683-0},
url={https://doi.org/10.1038/s41586-018-0683-0}
}

@Article{Krieg2021,
author={Krieg, Franziska
and Sercel, Peter C.
and Burian, Max
and Andrusiv, Hordii
and Bodnarchuk, Maryna I.
and Stöferle, Thilo
and Mahrt, Rainer F.
and Naumenko, Denys
and Amenitsch, Heinz
and Rainò, Gabriele
and Kovalenko, Maksym V.},
title={Monodisperse Long-Chain Sulfobetaine-Capped CsPbBr3 Nanocrystals and Their Superfluorescent Assemblies},
journal={ACS Central Science},
year={2021},
month={Jan},
day={27},
publisher={American Chemical Society},
volume={7},
number={1},
pages={135-144},
issn={2374-7943},
doi={10.1021/acscentsci.0c01153},
url={https://doi.org/10.1021/acscentsci.0c01153}
}

@article{Wang2021,
author = {Wang, Ya-Kun and Yuan, Fanglong and Dong, Yitong and Li, Jiao-Yang and Johnston, Andrew and Chen, Bin and Saidaminov, Makhsud I. and Zhou, Chun and Zheng, Xiaopeng and Hou, Yi and Bertens, Koen and Ebe, Hinako and Ma, Dongxin and Deng, Zhengtao and Yuan, Shuai and Chen, Rui and Sagar, Laxmi Kishore and Liu, Jiakai and Fan, James and Li, Peicheng and Li, Xiyan and Gao, Yuan and Fung, Man-Keung and Lu, Zheng-Hong and Bakr, Osman M. and Liao, Liang-Sheng and Sargent, Edward H.},
title = {All-Inorganic Quantum-Dot LEDs Based on a Phase-Stabilized О±-CsPbI3 Perovskite},
journal = {Angewandte Chemie International Edition},
volume = {60},
number = {29},
pages = {16164-16170},
doi = {https://doi.org/10.1002/anie.202104812},
url = {https://onlinelibrary.wiley.com/doi/abs/10.1002/anie.202104812},
eprint = {https://onlinelibrary.wiley.com/doi/pdf/10.1002/anie.202104812},
year = {2021}
}

@article{Moon2019,
author = {Moon, Hyungsuk and Lee, Changmin and Lee, Woosuk and Kim, Jungwoo and Chae, Heeyeop},
title = {Stability of Quantum Dots, Quantum Dot Films, and Quantum Dot Light-Emitting Diodes for Display Applications},
journal = {Advanced Materials},
volume = {31},
number = {34},
pages = {1804294},
doi = {https://doi.org/10.1002/adma.201804294},
url = {https://advanced.onlinelibrary.wiley.com/doi/abs/10.1002/adma.201804294},
eprint = {https://advanced.onlinelibrary.wiley.com/doi/pdf/10.1002/adma.201804294},
year = {2019}
}

@article{Jang2023,
author = {Jang, Eunjoo and Jang, Hyosook},
title = {Review: Quantum Dot Light-Emitting Diodes},
journal = {Chemical Reviews},
volume = {123},
number = {8},
pages = {4663-4692},
year = {2023},
doi = {10.1021/acs.chemrev.2c00695},
    note ={PMID: 36795794},

URL = {

        https://doi.org/10.1021/acs.chemrev.2c00695



},
eprint = {

        https://doi.org/10.1021/acs.chemrev.2c00695



}

}

@article{Chern2019,
doi = {10.1088/2050-6120/aaf6f8},
url = {https://doi.org/10.1088/2050-6120/aaf6f8},
year = {2019},
month = {jan},
publisher = {IOP Publishing},
volume = {7},
number = {1},
pages = {012005},
author = {Chern, Margaret and Kays, Joshua C and Bhuckory, Shashi and Dennis, Allison M},
title = {Sensing with photoluminescent semiconductor quantum dots},
journal = {Methods and Applications in Fluorescence}
}

@Article{Lesiak2019,
AUTHOR = {Lesiak, Anna and Drzozga, Kamila and Cabaj, Joanna and Bański, Mateusz and Malecha, Karol and Podhorodecki, Artur},
TITLE = {Optical Sensors Based on II-VI Quantum Dots},
JOURNAL = {Nanomaterials},
VOLUME = {9},
YEAR = {2019},
NUMBER = {2},
ARTICLE-NUMBER = {192},
PAGES= { },
URL = {https://www.mdpi.com/2079-4991/9/2/192},
PubMedID = {30717393},
ISSN = {2079-4991},
DOI = {10.3390/nano9020192}
}

@Article{Frasco2009,
AUTHOR = {Frasco, Manuela F. and Chaniotakis, Nikos},
TITLE = {Semiconductor Quantum Dots in Chemical Sensors and Biosensors},
JOURNAL = {Sensors},
VOLUME = {9},
YEAR = {2009},
NUMBER = {9},
PAGES = {7266--7286},
URL = {https://www.mdpi.com/1424-8220/9/9/7266},
PubMedID = {22423206},
ISSN = {1424-8220},
DOI = {10.3390/s90907266}
}

@article{galstyan2021,
  title={“Quantum dots: Perspectives in next-generation chemical gas sensors”--A review},
  author={Galstyan, Vardan},
  journal={Analytica Chimica Acta},
  volume={1152},
  pages={238192},
  year={2021},
  publisher={Elsevier}
}

@article{doNascimento2019,
  title={CdSe quantum dots as fluorescent nanomarkers for diesel oil},
  author={do Nascimento, Aquiles Silva and Cabral Filho, Paulo Euz{\'e}bio and Fontes, Adriana and Santos, Beate Saegesser and de Carvalho, Florival Rodrigues and Stragevitch, Luiz and Leite, Elisa Soares},
  journal={Fuel},
  volume={239},
  pages={1055--1060},
  year={2019},
  publisher={Elsevier}
}

@Article{Huang2024,
author ="Huang, Shiyu and Huang, Gangliang",
title  ="The utilization of quantum dot labeling as a burgeoning technique in the field of biological imaging",
journal  ="RSC Adv.",
year  ="2024",
volume  ="14",
issue  ="29",
pages  ="20884-20897",
publisher  ="The Royal Society of Chemistry",
doi  ="10.1039/D4RA04402A",
url  ="http://dx.doi.org/10.1039/D4RA04402A"}

@article{Thoulouli2008,
author = {Tholouli, E and Sweeney, E and Barrow, E and Clay, V and Hoyland, JA and Byers, RJ},
title = {Quantum dots light up pathology},
journal = {The Journal of Pathology},
volume = {216},
number = {3},
pages = {275-285},
doi = {https://doi.org/10.1002/path.2421},
url = {https://pathsocjournals.onlinelibrary.wiley.com/doi/abs/10.1002/path.2421},
eprint = {https://pathsocjournals.onlinelibrary.wiley.com/doi/pdf/10.1002/path.2421},
year = {2008}
}

@article{Jin2011,
author = {Jin, Shan and Hu, Yanxi and Gu, Zhanjun and Liu, Lei and Wu, Hai-Chen},
title = {Application of Quantum Dots in Biological Imaging},
journal = {Journal of Nanomaterials},
volume = {2011},
number = {1},
pages = {834139},
doi = {https://doi.org/10.1155/2011/834139},
url = {https://onlinelibrary.wiley.com/doi/abs/10.1155/2011/834139},
eprint = {https://onlinelibrary.wiley.com/doi/pdf/10.1155/2011/834139},
year = {2011}
}

@article{Ahmad2025,
author = {Ahmad, Imtiaz and Khan, Aziz and Rehman, Amna and Zahid, Maryam and Khalid, Jaweria and Saleem, Sidra and Majeed, Umer and Khan, Qasim and Maqbool, Muhammad},
year = {2025},
month = {05},
pages = {},
title = {Blinking effect in quantum dots, its suppression mechanism, and applications in medical imaging and biosensing: A review},
volume = {7},
journal = {AVS Quantum Science},
doi = {10.1116/5.0244906}
}

@article{Hao2024,
author = {Hao, Mengmeng and Ding, Shanshan and Gaznaghi, Sabah and Cheng, Huiyuan and Wang, Lianzhou},
title = {Perovskite Quantum Dot Solar Cells: Current Status and Future Outlook},
journal = {ACS Energy Letters},
volume = {9},
number = {1},
pages = {308-322},
year = {2024},
doi = {10.1021/acsenergylett.3c01983},

URL = {

        https://doi.org/10.1021/acsenergylett.3c01983



},
eprint = {

        https://doi.org/10.1021/acsenergylett.3c01983



}

}

@article{Semonin2012,
title = {Quantum dots for next-generation photovoltaics},
journal = {Materials Today},
volume = {15},
number = {11},
pages = {508-515},
year = {2012},
issn = {1369-7021},
doi = {https://doi.org/10.1016/S1369-7021(12)70220-1},
url = {https://www.sciencedirect.com/science/article/pii/S1369702112702201},
author = {Octavi E. Semonin and Joseph M. Luther and Matthew C. Beard}
}

@article{kovalenko2015,
  title={Opportunities and challenges for quantum dot photovoltaics},
  author={Kovalenko, Maksym V},
  journal={Nature Nanotechnology},
  volume={10},
  number={12},
  pages={994--997},
  year={2015},
  publisher={Nature Publishing Group UK London}
}

@article{Liu2022,
author = {Liu, Lu and Najar, Adel and Wang, Kai and Du, Minyong and Liu, Shengzhong (Frank)},
title = {Perovskite Quantum Dots in Solar Cells},
journal = {Advanced Science},
volume = {9},
number = {7},
pages = {2104577},
doi = {https://doi.org/10.1002/advs.202104577},
url = {https://advanced.onlinelibrary.wiley.com/doi/abs/10.1002/advs.202104577},
eprint = {https://advanced.onlinelibrary.wiley.com/doi/pdf/10.1002/advs.202104577},
year = {2022}
}

@article{Carey2015,
author = {Carey, Graham H. and Abdelhady, Ahmed L. and Ning, Zhijun and Thon, Susanna M. and Bakr, Osman M. and Sargent, Edward H.},
title = {Colloidal Quantum Dot Solar Cells},
journal = {Chemical Reviews},
volume = {115},
number = {23},
pages = {12732-12763},
year = {2015},
doi = {10.1021/acs.chemrev.5b00063},
    note ={PMID: 26106908},

URL = {

        https://doi.org/10.1021/acs.chemrev.5b00063



},
eprint = {

        https://doi.org/10.1021/acs.chemrev.5b00063



}

}

@article{Duan2021,
author = {Duan, Leiping and Hu, Long and Guan, Xinwei and Lin, Chun-Ho and Chu, Dewei and Huang, Shujuan and Liu, Xiaogang and Yuan, Jianyu and Wu, Tom},
title = {Quantum Dots for Photovoltaics: A Tale of Two Materials},
journal = {Advanced Energy Materials},
volume = {11},
number = {20},
pages = {2100354},
doi = {https://doi.org/10.1002/aenm.202100354},
url = {https://advanced.onlinelibrary.wiley.com/doi/abs/10.1002/aenm.202100354},
eprint = {https://advanced.onlinelibrary.wiley.com/doi/pdf/10.1002/aenm.202100354},
year = {2021}
}

@article{ryu2025,
author = {Ryu, Jehyeok and Krivenkov, Victor and Olejniczak, Adam and Nikitin, Alexey Y. and Rakovich, Yury},
    title = {Perovskite nanocrystals as emerging single-photon emitters: Progress, challenges, and opportunities},
    journal = {Applied Physics Reviews},
    volume = {12},
    number = {4},
    pages = {041323},
    year = {2025},
    month = {12},
    issn = {1931-9401},
    doi = {10.1063/5.0282667},
    url = {https://doi.org/10.1063/5.0282667},
    eprint = {https://pubs.aip.org/aip/apr/article-pdf/doi/10.1063/5.0282667/20832939/041323_1_5.0282667.pdf},
}

@article{senellart2017,
  title={High-performance semiconductor quantum-dot single-photon sources},
  author={Senellart, Pascale and Solomon, Glenn and White, Andrew},
  journal={Nature nanotechnology},
  volume={12},
  number={11},
  pages={1026--1039},
  year={2017},
  publisher={Nature Publishing Group UK London}
}

@article{arakawa2020,
  title={Progress in quantum-dot single photon sources for quantum information technologies: A broad spectrum overview},
  author={Arakawa, Yasuhiko and Holmes, Mark J},
  journal={Applied Physics Reviews},
  volume={7},
  number={2},
  pages={021309},
  year={2020},
  publisher={AIP Publishing}
}

@article{rakhlin2023,
  title={Demultiplexed single-photon source with a quantum dot coupled to microresonator},
  author={Rakhlin, MV and Galimov, AI and Dyakonov, IV and Skryabin, NN and Klimko, GV and Kulagina, MM and Zadiranov, Yu M and Sorokin, SV and Sedova, IV and Guseva, Yu A and others},
  journal={Journal of Luminescence},
  volume={253},
  pages={119496},
  year={2023},
  publisher={Elsevier}
}

@article{balitskii2024,
  title={Recent Developments in Halide Perovskite Nanocrystals for Indirect X-ray Detection},
  author={Balitskii, Olexiy and Sytnyk, Mykhailo and Heiss, Wolfgang},
  journal={Advanced Materials Technologies},
  volume={9},
  number={20},
  pages={2400150},
  year={2024},
  publisher={Wiley Online Library}
}

@article{YannHeng2025,
doi = {10.1088/1742-6596/2974/1/012026},
url = {https://doi.org/10.1088/1742-6596/2974/1/012026},
year = {2025},
month = {mar},
publisher = {IOP Publishing},
volume = {2974},
number = {1},
pages = {012026},
author = {Yann Heng, Han and Mohammad, Sabah M. and Alwani Zainuri, Dian and Mustaqim Rosli, Mohd and Razak Ibrahim, Abdul and Abdullah, Mundzir},
title = {Exploring the Optical and Nonlinear Optical Responses of Phenylammonium Bismuth Chloride Perovskite Quantum Dots},
journal = {Journal of Physics: Conference Series}
}

@article{heo2018,
  title={High-performance next-generation perovskite nanocrystal scintillator for nondestructive X-ray imaging},
  author={Heo, Jin Hyuck and Shin, Dong Hee and Park, Jin Kyoung and Kim, Do Hun and Lee, Sang Jin and Im, Sang Hyuk},
  journal={Advanced Materials},
  volume={30},
  number={40},
  pages={1801743},
  year={2018},
  publisher={Wiley Online Library}
}

@article{livache2019,
  title={A colloidal quantum dot infrared photodetector and its use for intraband detection},
  author={Livache, Cl{\'e}ment and Martinez, Bertille and Goubet, Nicolas and Gr{\'e}boval, Charlie and Qu, Junling and Chu, Audrey and Royer, S{\'e}bastien and Ithurria, Sandrine and Silly, Mathieu G and Dubertret, Benoit and others},
  journal={Nature communications},
  volume={10},
  number={1},
  pages={2125},
  year={2019},
  publisher={Nature Publishing Group UK London}
}

@article{zhukov2021,
  title={Quantum-dot microlasers based on whispering gallery mode resonators},
  author={Zhukov, AE and Kryzhanovskaya, NV and Moiseev, EI and Maximov, MV},
  journal={Light: Science \& Applications},
  volume={10},
  number={1},
  pages={80},
  year={2021},
  publisher={Nature Publishing Group UK London}
}

@article{park2021,
  title={Colloidal quantum dot lasers},
  author={Park, Young-Shin and Roh, Jeongkyun and Diroll, Benjamin T and Schaller, Richard D and Klimov, Victor I},
  journal={Nature Reviews Materials},
  volume={6},
  number={5},
  pages={382--401},
  year={2021},
  publisher={Nature Publishing Group UK London}
}

@article{zhang2021,
  title={Halide perovskite semiconductor lasers: materials, cavity design, and low threshold},
  author={Zhang, Qing and Shang, Qiuyu and Su, Rui and Do, T Thu Ha and Xiong, Qihua},
  journal={Nano Letters},
  volume={21},
  number={5},
  pages={1903--1914},
  year={2021},
  publisher={ACS Publications}
}

@article{yang2021,
  title={Efficient, stable, and photoluminescence intermittency-free CdSe-based quantum dots in the full-color range},
  author={Yang, Changgang and Xiao, Ruilin and Zhou, Sirong and Yang, Yonggang and Zhang, Guofeng and Li, Bin and Guo, Wenli and Han, Xue and Wang, Danhong and Bai, Xiuqing and others},
  journal={ACS Photonics},
  volume={8},
  number={8},
  pages={2538--2547},
  year={2021},
  publisher={ACS Publications}
}

@Article{Frantsuzov2008,
  author =      {P. A. Frantsuzov and M. Kuno and B. Jank\'o and R. A. Marcus},
  title =       {Universal emission intermittency in quantum dots, nanorods and nanowires},
  journal =     {Nat. Phys.},
  volume =      {4},
  number =      {},
  pages =       {519--522},
  year =        {2008}
}
\bibliographystyle{rsc}

\end{document}


\maketitle

\section*{Supplementary Note 1: Statistical properties of the relative quantum yield estimators}

\subsection*{S1.1. Distribution of the maximum of binned photon counts}
Consider a blinking trajectory consisting of $M_\textrm{tot}$ time bins. The number of photons detected in bin $j$, denoted $n_j$, is a Poisson-distributed random variable with mean $N_j$. The observed maximum intensity is
\begin{equation}
    n_\textrm{max} = \max_{j} n_j.
\end{equation}
We are interested in the case where the true maximum mean intensity $N_\textrm{max} = \max_j N_j$ is attained in $M_\textrm{max}$ bins. We assume that throughout the trajectory, $N_\textrm{max}$ is sufficiently large so that the Poisson distribution is well approximated by a normal distribution with mean $N_\textrm{max}$ and variance $N_\textrm{max}$. The problem then reduces to finding the distribution of the maximum of $M_\textrm{max}$ independent normal random variables.

According to the Fisher-Tippett-Gnedenko theorem (see, \textit{e.g.}, Refs.\citenum{gnedenko1992,leadbetter1983,deHaan2006}), the maximum of $M$ independent standard normal random variables converges, as $M \to \infty$, to the Gumbel distribution with the cumulative distribution function $F_G(g) = \exp(-e^{-g})$. The mean and variance of a Gumbel random variable $G$ are
\begin{equation}
    \mathbb{E}[G] = \gamma, \qquad \mathbb{V}\textrm{ar}[G] = \frac{\pi^2}{6},
\end{equation}
where $\gamma$ is the Euler-Mascheroni constant. The maximum of $M$ standard normal variables $Z_i$ can be expressed in terms of $G$ as
\begin{equation}
    \max_{i=1,\dots,M} Z_i \approx a_M + b_M G,
\end{equation}
with the scaling constants\cite{leadbetter1983}
\begin{equation}
    a_M = \sqrt{2\ln M}- \frac{\ln(4\pi\ln M)}{2\sqrt{2\ln M}}, \qquad b_M = \frac{1}{\sqrt{2\ln M}}.
\end{equation}
Taking the expectation yields
\begin{equation}
    \mathbb{E}\left[\max_i Z_i\right] = a_M + \gamma \, b_M = \sqrt{2\ln M} + \frac{\gamma}{\sqrt{2\ln M}}- \frac{\ln(4\pi\ln M)}{2\sqrt{2\ln M}} + o\left(\frac{1}{\sqrt{\ln M}}\right).
\end{equation}
Thus, for a general normal variable with mean $\mu$ and variance $\sigma^2$, the expected maximum is $\mu + \sigma \, \mathbb{E}[\max_i Z_i]$, and we identify
\begin{equation}
    K(M) = \sqrt{2\ln M} + \frac{\gamma}{\sqrt{2\ln M}}- \frac{\ln(4\pi\ln M)}{2\sqrt{2\ln M}} + o\left(\frac{1}{\sqrt{\ln M}}\right).
    \label{eqn:S_K_exact}
\end{equation}
The leading-order approximation $K(M) \approx \sqrt{2\ln M}$ is sufficient for most practical purposes, the higher-order terms are given here for completeness. In all simulations and numerical calculations presented in this work, the full expression\eqref{eqn:S_K_exact} was used.

Applying this result to our case with $\mu = N_\textrm{max}$ and $\sigma = \sqrt{N_\textrm{max}}$, we obtain
\begin{equation}
    \mathbb{E}[n_\textrm{max}] = N_\textrm{max} + K(M_\textrm{max}) \sqrt{N_\textrm{max}} \equiv \mathcal{N}_\textrm{max}.
    \label{eqn:S_Nmax_def}
\end{equation}
The variance of the maximum follows from $\mathbb{V}\textrm{ar}[G] = \pi^2/6$ and the scaling $b_M = 1/\sqrt{2\ln M}$:
\begin{equation}
    \mathbb{V}\textrm{ar}[n_\textrm{max}] = \sigma^2 \, b_M^2 \, \mathbb{V}\textrm{ar}[G] = N_\textrm{max} \cdot \frac{1}{2\ln M_\textrm{max}} \cdot \frac{\pi^2}{6} = \frac{N_\textrm{max} \pi^2}{12 \ln M_\textrm{max}}.
    \label{eqn:S_variance_Imax}
\end{equation}

\subsection*{S1.2. The naive estimator and its bias}

The naive estimator of the relative quantum yield is defined in the main text as
\begin{equation}
    \langle \hat{Q} \rangle^{(0)} = \frac{\langle n \rangle}{n_\textrm{max}},
    \label{eqn:S_Q0_def}
\end{equation}
where $\langle n \rangle = M_\textrm{tot}^{-1} \sum_j n_j$ is the mean photon count per bin. The true relative quantum yield, which this estimator aims to recover, is
\begin{equation}
    \langle Q \rangle_\textrm{true} = \frac{\langle N \rangle}{N_\textrm{max}},
    \label{eqn:S_Qtrue_def}
\end{equation}
where $\langle N \rangle = M_\textrm{tot}^{-1} \sum_j N_j$ is the true mean intensity and $N_\textrm{max} = \max_j N_j$ is the true maximum mean intensity.

Both the numerator and the denominator in Eq. \eqref{eqn:S_Q0_def} are random variables. We evaluate the expectation by expanding the ratio around $(\langle N \rangle, \mathcal{N}_\textrm{max})$. To second order in the fluctuations,
\begin{equation}
    \mathbb{E}\left[\frac{\langle n \rangle}{n_\textrm{max}}\right] \approx \frac{\langle N \rangle}{\mathcal{N}_\textrm{max}} + \frac{\langle N \rangle}{\mathcal{N}_\textrm{max}^3} \mathbb{V}\textrm{ar}[n_\textrm{max}]- \frac{1}{\mathcal{N}_\textrm{max}^2} \mathbb{C}\textrm{ov}\left[\langle n \rangle, n_\textrm{max}\right].
\end{equation}
The linear terms vanish because $\mathbb{E}[\langle n \rangle- \langle N \rangle] = 0$ and $\mathbb{E}[n_\textrm{max}- \mathcal{N}_\textrm{max}] = 0$.

The covariance term can be evaluated explicitly. Writing $\langle n \rangle = M_\textrm{tot}^{-1} \sum_i n_i$, we have
\begin{equation}
    \mathbb{C}\textrm{ov}\left[\langle n \rangle, n_\textrm{max}\right] = \frac{1}{M_\textrm{tot}} \sum_{i=1}^{M_\textrm{tot}} \mathbb{C}\textrm{ov}\left[n_i, n_\textrm{max}\right].
\end{equation}
The covariance $\mathbb{C}\textrm{ov}[n_i, n_\textrm{max}]$ is non-negligible only for those bins where $n_i$ is close to $n_\textrm{max}$, \textit{i.e.}, for bins belonging to the set of $M_\textrm{max}$ bins with the highest true mean intensity $N_\textrm{max}$. For all other bins, $n_i$ and $n_\textrm{max}$ are essentially independent, and their covariance vanishes. Therefore,
\begin{equation}
    \sum_{i=1}^{M_\textrm{tot}} \mathbb{C}\textrm{ov}\left[n_i, n_\textrm{max}\right] \approx M_\textrm{max} \cdot \mathbb{C}\textrm{ov}\left[n_{i^*}, n_\textrm{max}\right] \approx M_\textrm{max} \cdot \mathbb{V}\textrm{ar}[n_\textrm{max}],
\end{equation}
where $i^*$ denotes a bin attaining the maximum, and we used the fact that for such bins $n_{i^*}$ and $n_\textrm{max}$ are strongly correlated, with $\mathbb{C}\textrm{ov}[n_{i^*}, n_\textrm{max}] \approx \mathbb{V}\textrm{ar}[n_\textrm{max}]$. It follows that
\begin{equation}
    \mathbb{C}\textrm{ov}\left[\langle n \rangle, n_\textrm{max}\right] \approx \frac{M_\textrm{max}}{M_\textrm{tot}} \, \mathbb{V}\textrm{ar}[n_\textrm{max}] \ll \mathbb{V}\textrm{ar}[n_\textrm{max}],
\end{equation}
since $M_\textrm{max} \ll M_\textrm{tot}$ in any realistic blinking trajectory. The covariance term in the expansion is therefore suppressed by the factor $M_\textrm{max}/M_\textrm{tot}$ and can be safely neglected.

Dropping the covariance term and using the definitions \eqref{eqn:S_Nmax_def} and \eqref{eqn:S_Qtrue_def}, we obtain
\begin{equation}
    \begin{aligned}
        \mathbb{E}\left[\langle \hat{Q} \rangle^{(0)}\right] &\approx \frac{\langle N \rangle}{N_\textrm{max} + K\sqrt{N_\textrm{max}}} + \frac{\langle N \rangle}{\mathcal{N}_\textrm{max}^3} \mathbb{V}\textrm{ar}[n_\textrm{max}] \\
        &\approx \langle Q \rangle_\textrm{true} \left(1- K\frac{\sqrt{N_\textrm{max}}}{N_\textrm{max}}\right) + \mathcal{O}\left(\frac{1}{N_\textrm{max}\ln M_\textrm{max}}\right),
    \end{aligned}
\end{equation}
where we expanded $1/(N_\textrm{max} + K\sqrt{N_\textrm{max}})$ to leading order in $K/\sqrt{N_\textrm{max}}$. The bias is therefore
\begin{equation}
    \textrm{Bias}\left[\langle \hat{Q} \rangle^{(0)}\right] = \mathbb{E}\left[\langle \hat{Q} \rangle^{(0)}\right]- \langle Q \rangle_\textrm{true} \approx-\langle Q \rangle_\textrm{true} \, K(M_\textrm{max}) \, \frac{\sqrt{N_\textrm{max}}}{N_\textrm{max}}.
    \label{eqn:S_bias_Q0}
\end{equation}
The bias is negative: the naive estimator systematically underestimates the true relative quantum yield.

\subsection*{S1.3. The self-consistent estimator and its bias}

To correct the bias of the naive estimator, we introduce the self-consistent estimate of $N_\textrm{max}$. Setting $n_\textrm{max} = \mathbb{E}[n_\textrm{max}]$ in Eq. \eqref{eqn:S_Nmax_def} and solving for $N_\textrm{max}$ yields
\begin{equation}
    \hat{N}^{(s)}_\textrm{max} = \frac{1}{4}\left(\sqrt{K^2 + 4 n_\textrm{max}}- K\right)^2,
    \label{eqn:S_Nmax_hat}
\end{equation}
with $K = K(M_\textrm{max})$. The self-consistent estimator of the relative quantum yield is then
\begin{equation}
    \langle \hat{Q} \rangle^{(s)} = \frac{\langle n \rangle}{\hat{N}^{(s)}_\textrm{max}}.
    \label{eqn:S_Qs_def}
\end{equation}

As before, both the numerator and the denominator are random variables. We evaluate the expectation by expanding the ratio around $(\langle N \rangle, \mathcal{N}_\textrm{max})$. It is convenient to define the function $f(x) = \frac{1}{4}(\sqrt{K^2 + 4x}- K)^2$, so that $\hat{N}^{(s)}_\textrm{max} = f(n_\textrm{max})$ and $N_\textrm{max} = f(\mathcal{N}_\textrm{max})$. Expanding the ratio $\langle n \rangle / f(n_\textrm{max})$ to second order in the fluctuations gives
\begin{equation}
    \mathbb{E}\left[\frac{\langle n \rangle}{f(n_\textrm{max})}\right] \approx \frac{\langle N \rangle}{N_\textrm{max}} + \frac{\langle N \rangle}{2} \, g''(\mathcal{N}_\textrm{max}) \, \mathbb{V}\textrm{ar}[n_\textrm{max}]- \frac{f'(\mathcal{N}_\textrm{max})}{N_\textrm{max}^2} \, \mathbb{C}\textrm{ov}\left[\langle n \rangle, n_\textrm{max}\right],
    \label{eqn:S_expand_Qs}
\end{equation}
where $g(x) = 1/f(x)$ and we used $g(\mathcal{N}_\textrm{max}) = 1/N_\textrm{max}$. The linear terms vanish because $\mathbb{E}[\langle n \rangle- \langle N \rangle] = 0$ and $\mathbb{E}[n_\textrm{max}- \mathcal{N}_\textrm{max}] = 0$.

The covariance term is the same as that evaluated in the previous subsection. Writing $\langle n \rangle = M_\textrm{tot}^{-1} \sum_i n_i$, we have
\begin{equation}
    \mathbb{C}\textrm{ov}\left[\langle n \rangle, n_\textrm{max}\right] = \frac{1}{M_\textrm{tot}} \sum_{i=1}^{M_\textrm{tot}} \mathbb{C}\textrm{ov}\left[n_i, n_\textrm{max}\right].
\end{equation}
The sum receives non-negligible contributions only from the $M_\textrm{max}$ bins with the highest true mean intensity, where $n_i$ is strongly correlated with $n_\textrm{max}$. For these bins, $\mathbb{C}\textrm{ov}[n_{i^*}, n_\textrm{max}] \approx \mathbb{V}\textrm{ar}[n_\textrm{max}]$, yielding
\begin{equation}
    \mathbb{C}\textrm{ov}\left[\langle n \rangle, n_\textrm{max}\right] \approx \frac{M_\textrm{max}}{M_\textrm{tot}} \, \mathbb{V}\textrm{ar}[n_\textrm{max}] \ll \mathbb{V}\textrm{ar}[n_\textrm{max}],
\end{equation}
since $M_\textrm{max} \ll M_\textrm{tot}$. The covariance term in Eq. \eqref{eqn:S_expand_Qs} is therefore suppressed by $M_\textrm{max}/M_\textrm{tot}$ and can be dropped.

The remaining second-order term requires the derivatives of $f(x)$. A straightforward calculation gives
\begin{equation}
    f'(\mathcal{N}_\textrm{max}) = \frac{1}{1 + K/(2\sqrt{N_\textrm{max}})}, \qquad f''(\mathcal{N}_\textrm{max}) = \frac{2K}{(K + 2\sqrt{N_\textrm{max}})^3},
\end{equation}
and
\begin{equation}
    g''(\mathcal{N}_\textrm{max}) = \frac{2[f'(\mathcal{N}_\textrm{max})]^2- f''(\mathcal{N}_\textrm{max}) N_\textrm{max}}{N_\textrm{max}^3}.
\end{equation}
Substituting the expressions for $f'$ and $f''$ and using $\mathbb{V}\textrm{ar}[n_\textrm{max}]$ from Eq. \eqref{eqn:S_variance_Imax}, one obtains, after straightforward algebra,
\begin{equation}
    \mathbb{E}\left[\langle \hat{Q} \rangle^{(s)}\right] \approx \langle Q \rangle_\textrm{true} + \frac{\pi^2 \langle Q \rangle_\textrm{true}}{12 N_\textrm{max} \ln M_\textrm{max}}.
    \label{eqn:S_bias_Qs}
\end{equation}
The absolute value of the bias,
\begin{equation}
    \left|\textrm{Bias}\left[\langle \hat{Q} \rangle^{(s)}\right]\right| \approx \frac{\pi^2 \langle Q \rangle_\textrm{true}}{12 N_\textrm{max} \ln M_\textrm{max}},
    \label{eqn:S_bias_Qs_abs}
\end{equation}
is much smaller than $\langle Q \rangle_\textrm{true}$ for all realistic parameters. Comparing with Eq. \eqref{eqn:S_bias_Q0}, we obtain the ratio of biases:
\begin{equation}
    \left|\frac{\textrm{Bias}\left[\langle \hat{Q} \rangle^{(s)}\right]}{\textrm{Bias}\left[\langle \hat{Q} \rangle^{(0)}\right]}\right| = \frac{\pi^2}{12 \ln M_\textrm{max} \, K(M_\textrm{max}) \, N_\textrm{max}^{1/2}} \ll 1.
    \label{eqn:S_bias_ratio}
\end{equation}

\subsection*{S1.4. Variance and mean-squared error}

We now evaluate the variances of the two estimators. For the naive estimator $\langle \hat{Q} \rangle^{(0)} = \langle n \rangle / n_\textrm{max}$, we again expand the ratio around $(\langle N \rangle, \mathcal{N}_\textrm{max})$. To leading order in the fluctuations,
\begin{equation}
    \langle \hat{Q} \rangle^{(0)}- \langle Q \rangle_\textrm{true} \approx \frac{1}{\mathcal{N}_\textrm{max}} (\langle n \rangle- \langle N \rangle)- \frac{\langle N \rangle}{\mathcal{N}_\textrm{max}^2} (n_\textrm{max}- \mathcal{N}_\textrm{max}).
\end{equation}
Squaring and taking the expectation yields
\begin{equation}
    \mathbb{V}\textrm{ar}\left[\langle \hat{Q} \rangle^{(0)}\right] \approx \frac{1}{\mathcal{N}_\textrm{max}^2} \mathbb{V}\textrm{ar}[\langle n \rangle] + \frac{\langle N \rangle^2}{\mathcal{N}_\textrm{max}^4} \mathbb{V}\textrm{ar}[n_\textrm{max}]- 2\frac{\langle N \rangle}{\mathcal{N}_\textrm{max}^3} \mathbb{C}\textrm{ov}\left[\langle n \rangle, n_\textrm{max}\right].
\end{equation}
The first term involves $\mathbb{V}\textrm{ar}[\langle n \rangle] = M_\textrm{tot}^{-1} \mathbb{V}\textrm{ar}[n_j] \sim M_\textrm{tot}^{-1} N_\textrm{max}$, which is negligible for large $M_\textrm{tot}$. The covariance term, as shown in the previous subsections, is suppressed by $M_\textrm{max}/M_\textrm{tot}$ and can also be dropped. The dominant contribution comes from the variance of $n_\textrm{max}$. Using $\mathbb{V}\textrm{ar}[n_\textrm{max}]$ from Eq. \eqref{eqn:S_variance_Imax} and $\langle N \rangle = \langle Q \rangle_\textrm{true} N_\textrm{max}$, we obtain
\begin{equation}
    \mathbb{V}\textrm{ar}\left[\langle \hat{Q} \rangle^{(0)}\right] \approx \frac{\langle Q \rangle_\textrm{true}^2 N_\textrm{max}^2}{\mathcal{N}_\textrm{max}^4} \cdot \frac{N_\textrm{max} \pi^2}{12 \ln M_\textrm{max}} = \frac{\langle Q \rangle_\textrm{true}^2 \pi^2}{12 \ln M_\textrm{max}} \cdot \frac{N_\textrm{max}^3}{\mathcal{N}_\textrm{max}^4}.
    \label{eqn:S_var_Q0}
\end{equation}

For the self-consistent estimator $\langle \hat{Q} \rangle^{(s)} = \langle n \rangle / f(n_\textrm{max})$, the same leading-order expansion gives
\begin{equation}
    \langle \hat{Q} \rangle^{(s)}- \langle Q \rangle_\textrm{true} \approx \frac{1}{N_\textrm{max}} (\langle n \rangle- \langle N \rangle)- \frac{\langle N \rangle f'(\mathcal{N}_\textrm{max})}{N_\textrm{max}^2} (n_\textrm{max}- \mathcal{N}_\textrm{max}).
\end{equation}
Squaring, taking the expectation, and dropping the terms involving $\mathbb{V}\textrm{ar}[\langle n \rangle]$ and the covariance (both negligible for the same reasons as above), we obtain
\begin{equation}
    \mathbb{V}\textrm{ar}\left[\langle \hat{Q} \rangle^{(s)}\right] \approx \frac{\langle N \rangle^2 [f'(\mathcal{N}_\textrm{max})]^2}{N_\textrm{max}^4} \, \mathbb{V}\textrm{ar}[n_\textrm{max}].
\end{equation}
For typical experimental parameters, $N_\textrm{max} \gg K$, so that $f'(\mathcal{N}_\textrm{max}) \approx 1$ [see Eq. \eqref{eqn:S_Nmax_hat} and the expression for $f'$]. Substituting $\mathbb{V}\textrm{ar}[n_\textrm{max}]$ and $\langle N \rangle = \langle Q \rangle_\textrm{true} N_\textrm{max}$ yields
\begin{equation}
    \mathbb{V}\textrm{ar}\left[\langle \hat{Q} \rangle^{(s)}\right] \approx \frac{\langle Q \rangle_\textrm{true}^2 \pi^2}{12 N_\textrm{max} \ln M_\textrm{max}}.
    \label{eqn:S_var_Qs}
\end{equation}

The mean-squared error (MSE) of an estimator is the sum of its variance and the square of its bias. From the results above, the following ordering holds for all $M_\textrm{max} > 10$:
\begin{equation}
    \left(\textrm{Bias}\left[\langle \hat{Q} \rangle^{(s)}\right]\right)^2 \ll \mathbb{V}\textrm{ar}\left[\langle \hat{Q} \rangle^{(0)}\right] < \mathbb{V}\textrm{ar}\left[\langle \hat{Q} \rangle^{(s)}\right] \ll \left(\textrm{Bias}\left[\langle \hat{Q} \rangle^{(0)}\right]\right)^2.
    \label{eqn:S_ordering}
\end{equation}
Consequently,
\begin{equation}
    \textrm{MSE}\left[\langle \hat{Q} \rangle^{(0)}\right] \approx \left(\textrm{Bias}\left[\langle \hat{Q} \rangle^{(0)}\right]\right)^2 \gg \textrm{MSE}\left[\langle \hat{Q} \rangle^{(s)}\right] \approx \mathbb{V}\textrm{ar}\left[\langle \hat{Q} \rangle^{(s)}\right].
    \label{eqn:S_MSE}
\end{equation}
Thus, the naive estimator is dominated by its systematic bias, whereas the error of the self-consistent estimator is governed by its variance. For $M_\textrm{max} > 10$, the self-consistent estimator yields a substantially smaller mean-squared error and is therefore the preferred choice.

The condition $M_\textrm{max} > 10$ for the dominance of the self-consistent estimator follows from a simple order-of-magnitude estimate. For a well-resolved blinking trajectory, $N_\textrm{max}$ is typically several tens of counts per bin (we use $N_\textrm{max} \approx 50$ as a representative value). When $N_\textrm{max}$ is too low, the normal approximation to the Poisson distribution and the extreme-value asymptotics become less reliable. With $N_\textrm{max} \approx 50$, the factor $K(M_\textrm{max}) = \sqrt{2\ln M_\textrm{max}}$ is of order $2$-$3$ for $M_\textrm{max}$ in the range $10$-$10^2$, and the bias ratio in Eq. \eqref{eqn:S_bias_ratio} is already below $3 \times 10^{-2}$. The mean-squared error of the self-consistent estimator is then over an order of magnitude smaller than that of the naive estimator. For smaller $M_\textrm{max}$, the extreme-value approximation itself becomes unreliable, as the Fisher-Tippett-Gnedenko limit requires a sufficiently large number of samples. In practice, $M_\textrm{max} > 10$ and $N_\textrm{max}$ of several tens of counts are typical for blinking trajectories longer than a few hundred bins, making the self-consistent estimator applicable to virtually all realistic experimental data sets.

\section*{Supplementary Note 2: Details of the blinking simulations}

\subsection*{S2.1. Two-state blinking: telegraph process algorithm}

The first simulation method models two-state A-type blinking as a stochastic telegraph process. The algorithm proceeds as follows.

The ON and OFF states are characterized by the pairs $\Gamma_\textrm{ON}$, $N_\textrm{ON}$ and $\Gamma_\textrm{OFF}$, $N_\textrm{OFF}$, chosen to satisfy Eq. 3 of the main text. The fraction of time spent in the OFF state is set by the parameter $p_\textrm{OFF} = T_\textrm{OFF} / (T_\textrm{ON} + T_\textrm{OFF})$. The characteristic switching time between states, $T_\textrm{switch}$, is taken to be the same for both directions. Starting from the ON state, a sequence of switching points is generated. The time to the next switching point is drawn from an exponential distribution with mean $T_\textrm{switch}$:
\begin{equation}
    \delta t =-T_\textrm{switch} \ln X,
    \label{eqn:S_tswitch}
\end{equation}
where $X$ is a random variable uniformly distributed on $[0,1]$. At each switching point, a new state is assigned by the following rule. If the current state is ON, the state switches to OFF with probability $p_\textrm{OFF}$ and remains ON with probability $1- p_\textrm{OFF}$. If the current state is OFF, the state switches to ON with probability $1- p_\textrm{OFF}$ and remains OFF with probability $p_\textrm{OFF}$. This procedure ensures that the trajectory spends a stationary fraction $p_\textrm{OFF}$ of time in the OFF state, while the times between successive switching points are exponentially distributed with mean $T_\textrm{switch}$.

Between switching points, the intensity and decay rate are assumed constant at their respective ON or OFF values. The continuous-time signal is then binned with a chosen bin size $\Delta t$. If one or more switching events fall within a single bin, the bin intensity is taken as the time-averaged value over that bin. The corresponding decay rate for the bin is obtained by weighting the ON- and OFF-state rates by their respective intensity contributions within the bin:
\begin{equation}
    \Gamma_j = \frac{N_\textrm{ON} \, t_\textrm{ON} \, \Gamma_\textrm{ON} + N_\textrm{OFF} \, t_\textrm{OFF} \, \Gamma_\textrm{OFF}}{N_\textrm{ON} \, t_\textrm{ON} + N_\textrm{OFF} \, t_\textrm{OFF}},
    \label{eqn:S_Gamma_weighted}
\end{equation}
where $t_\textrm{ON}$ and $t_\textrm{OFF}$ are the total dwell times of the respective states within the bin. This ensures that the brighter state contributes proportionally more to the effective $\Gamma_j$, as would be the case in a real FLID-based analysis. If no switching occurs within a bin, both quantities are taken from the current state.

After constructing the noise-free trajectories $N_j$ and $\Gamma_j$, experimental uncertainties are introduced. Poisson noise is added to the binned intensities:
\begin{equation}
    n_j \sim \textrm{Pois}(N_j).
\end{equation}
For the decay rate, we account for the estimation uncertainty arising from finite photon statistics. For each $\Gamma_j$, a random sample $\varepsilon$ is drawn from the standard normal distribution $\mathcal{N}(0,1)$, multiplied by $\Gamma_j / \sqrt{N_\textrm{fit}}$, and added to the true value:
\begin{equation}
    \hat{\Gamma}_j = \Gamma_j + \varepsilon \, \frac{\Gamma_j}{\sqrt{N_\textrm{fit}}}, \qquad \varepsilon \sim \mathcal{N}(0,1).
\end{equation}
Here $N_\textrm{fit}$ is the typical number of photons used in a maximum-likelihood fit, we set $N_\textrm{fit} = 1000$ throughout. We note that the $1/\sqrt{N_\textrm{fit}}$ scaling was originally derived for the variance of lifetime estimates in Ref. \citenum{Podshivaylov2023}. The analogous form for the decay rate follows directly. This procedure yields the simulated experimental trajectories $\{n_j\}$ and $\{\hat{\Gamma}_j\}$.

The HC mechanism is simulated using the same framework, with the simplification that the decay rate remains unchanged upon switching ($\Gamma_\textrm{ON} = \Gamma_\textrm{OFF} = \Gamma$). Only the intensity fluctuates between $N_\textrm{ON}$ and $N_\textrm{OFF}$. The same Poisson noise and $\Gamma$-broadening are applied.

\subsection*{S2.2. Multilevel TM blinking: TLS-based model}

The second simulation method implements the multilevel TM blinking model proposed in Ref. \citenum{Podshivaylov2023}. The model assumes that the quantum dot structure evolves over time due to a discrete set of independent stochastic two-level systems (TLSs). The switching of each TLS modifies the Huang-Rhys parameter $S$ for the nonradiative transition as
\begin{equation}
    S(t) = s_0 + \sum_i \sigma_i(t) s_i,
    \label{eqn:S_HR}
\end{equation}
where $\sigma_i(t) \in \{0,1\}$ describes the state of the $i$-th TLS, $s_0$ is the baseline Huang-Rhys parameter with all TLSs off, and $s_i$ is the contribution of the $i$-th TLS when active. According to Marcus-Jortner theory, the nonradiative capture rate depends on $S(t)$ as
\begin{equation}
    k_\textrm{nr}(t) = k_0 S^{\alpha}(t),
    \label{eqn:S_k_nr}
\end{equation}
where $k_0$ is a pre-exponential factor and $\alpha$ is the effective number of phonons involved in the transition. The switching times of individual TLSs are drawn from a log-uniform distribution spanning a broad range of time scales. Given the time sequence $S(t)$, the decay rate $\Gamma(t)$ and the mean bin intensity $N_j$ follow from the relations for TM blinking [Eq. 4 of the main text].

The simulation proceeds by generating a switching time for each TLS from its exponential distribution, in analogy with Eq. \eqref{eqn:S_tswitch}. The smallest of the obtained times is selected as the next switching event. The state of the corresponding TLS is flipped, and its next switching time is regenerated. For the remaining TLSs, instead of regenerating all switching times anew, the chosen switching time is subtracted from their previously generated times:
\begin{equation}
    T_i \leftarrow T_i- T_\textrm{switch}, \quad i \neq i^*,
\end{equation}
where $i^*$ is the index of the TLS that switched. This procedure is repeated until the required number of switching events is reached. For each switching event, the Huang-Rhys parameter $S$ is updated according to Eq. \eqref{eqn:S_HR}, and the corresponding $k_\textrm{nr}$, $\Gamma$, and $N$ are computed. Poisson noise and $\Gamma$-broadening are then applied as described in previous subsection.

This approach ensures that even TLSs with the longest switching times eventually switch, allowing fully developed blinking to be observed. The simulated trajectories reproduce all characteristic features of TM blinking: the photon-number distribution, the linear FLID, the ON/OFF time distributions, the long-time autocorrelation function, and the $1/f$-shaped power spectral density.

\subsection*{S2.3. Simulation parameters for Figure 2}

Figure 2 of the main text shows a simulated trajectory for A-type blinking. The parameters used are as follows. The radiative recombination rate is $k_\textrm{r} = 1/30 \textrm{ns}^{-1}$, and all nonradiative rates are set to zero: $k_\textrm{nr} = 0$, $k_\textrm{nr}' = 0$. The Auger rate is $k_\textrm{A} = 10 \, k_\textrm{r}$. The switching times in both directions are equal: $T^{+} = T^{-} = 0.03$ s, corresponding to $p_\textrm{OFF} = 0.5$. The maximum mean intensity is $N_\textrm{max} = 120$ counts per bin. The simulation was run for $10^4$ switching events with a bin size of $0.01$ s.

\subsection*{S2.4. Simulation parameters for Figure 3}

Figure 3 of the main text presents a simulated TM blinking trajectory based on the model of Ref. \cite{Podshivaylov2023}, using parameters obtained from a fit to experimental data of a single CdSeS/ZnS core/shell quantum dot (QD 3.2 in that reference). The radiative recombination rate is $k_\textrm{r} = 1/30 \textrm{ns}^{-1} \approx 3.33 \times 10^7 \textrm{s}^{-1}$. The nonradiative rate is given by $k_\textrm{nr}(t) = k_0 S^{\alpha}(t)$, with $k_0 / k_\textrm{r} = 6.17 \times 10^7$ and $\alpha = 10$. The Huang--Rhys parameter $S(t)$ is modulated by seven independent two-level systems, $S(t) = s_0 + \sum_{i=1}^7 \sigma_i(t) s_i$, with the baseline and individual contributions given in Table \ref{tab:S1}.

\begin{table}[h]
\centering
\begin{tabular}{c|c|c|c}
\hline
TLS index $i$ & $s_i$ & $p_i$ & $\Gamma_i$ ($\textrm{s}^{-1}$) \\
\hline
-- & $s_0 = 0.0756$ & -- & -- \\
1 & $0.0083$ & $0.397$ & $1.99 \times 10^{-2}$ \\
2 & $0.134$ & $0.237$ & $7.84 \times 10^{-2}$ \\
3 & $0.0349$ & $0.149$ & $3.08 \times 10^{-1}$ \\
4 & $0.0180$ & $0.347$ & $1.21$ \\
5 & $0.0590$ & $0.834$ & $4.76$ \\
6 & $0.0169$ & $0.305$ & $7.36 \times 10^{1}$ \\
7 & $0.0700$ & $0.0125$ & $1.14 \times 10^{3}$ \\
\hline
\end{tabular}
\caption{Parameters of the two-level systems modulating the Huang--Rhys parameter for QD 3.2 from Ref. \cite{Podshivaylov2023}.}
\label{tab:S1}
\end{table}

In Table \ref{tab:S1}, $p_i$ denotes the stationary probability that the $i$-th TLS is in the active state ($\sigma_i = 1$), and $\Gamma_i$ is the switching rate of that TLS (the inverse of its characteristic switching time). The mean intensity is $N_\textrm{med} = 88.2$ counts per bin. The simulation was run for $5 \times 10^4$ switching events with a bin size of $10$ ms.

\subsection*{S2.5. Reproduced blinking characteristics}

The simulations based on the multilevel TM model quantitatively preserve the known properties of all characteristic blinking observables. Representative examples obtained from the simulation of Fig. 3 are shown in Figure S1.

Panel (a) displays the ON-time (left) and OFF-time (right) probability distributions, both of which follow power-law decays over several decades. The red lines are guides to the eye indicating the power-law behaviour. Superimposed across both distributions is the intensity trajectory with a threshold (solid line) used to define the ON and OFF states. Panel (b) displays the long-time autocorrelation function $g^{(2)}(t)$ of the intensity trajectory. Panel (c) shows the power spectral density computed using modified Welch's method, following the procedure described in Ref. \citenum{Seth2021}. The PSD exhibits a $1/f^{1.3}$ dependence (red guide to the eye), consistent with the power-law expectation. Together, these diagnostics confirm that the simulation faithfully reproduces all established statistical fingerprints of TM blinking.

\begin{figure}[h]
\centering
\includegraphics[width=0.99\linewidth]{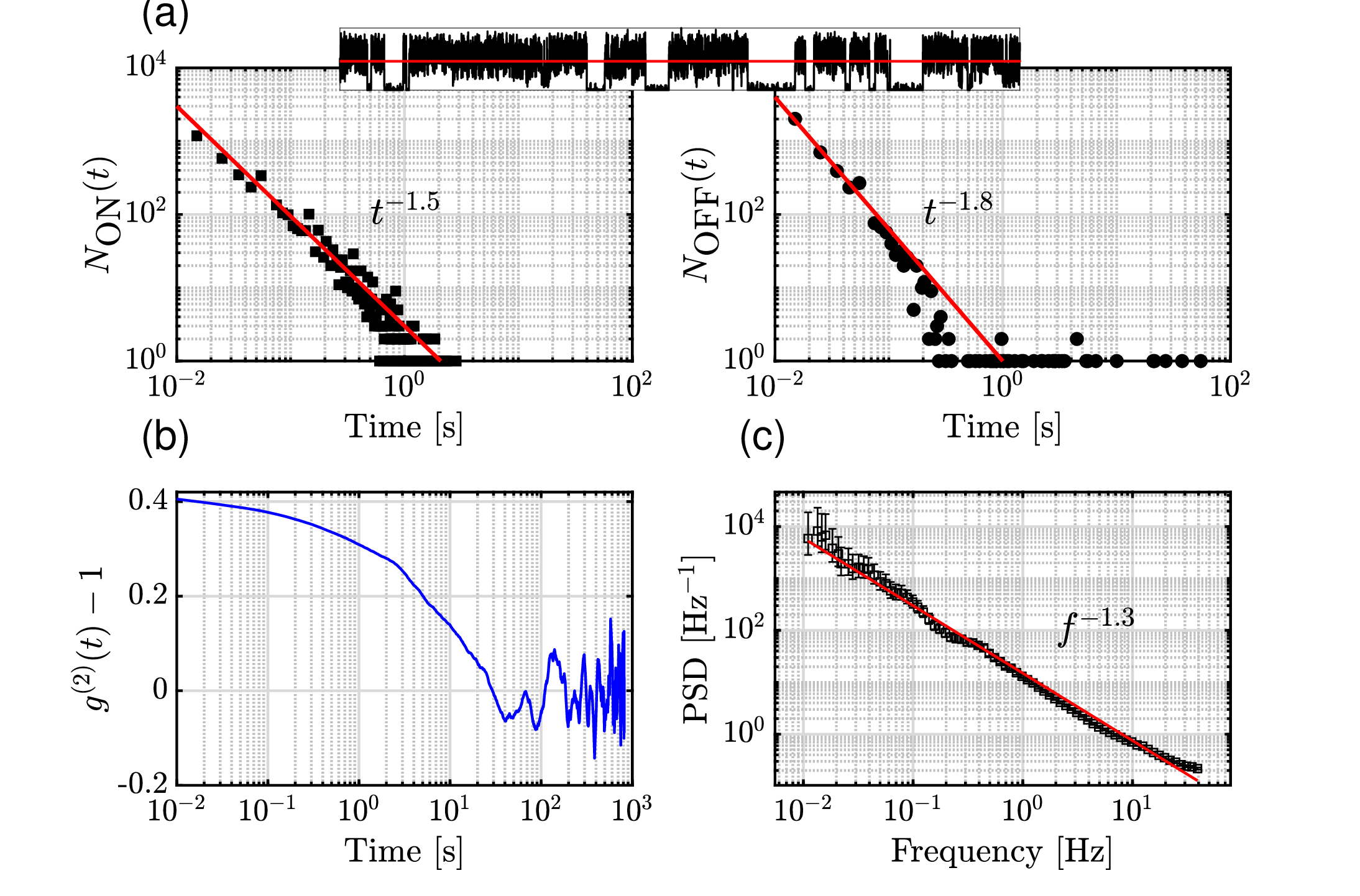}
\caption{Statistical diagnostics of the simulated TM blinking trajectory from Fig. 3. (a) Probability distributions of ON times (left) and OFF times (right), with red lines indicating power-law behaviour. The intensity trajectory with a threshold (solid line) defining ON and OFF states is superimposed across both panels. (b) Autocorrelation function $g^{(2)}(\tau)$. (c) Power spectral density computed via modified Welch's method, with a red line showing a $1/f^{1.3}$ guide to the eye.}
\label{fig:S1}
\end{figure}

\section*{Supplementary Note 3: Simulation parameters for the estimator comparison and the 2D criterion}

\subsection*{S3.1. Parameters for Figure 4: comparison of estimators}

Figure 4 of the main text presents a systematic comparison of the naive estimator $\langle\hat{Q}\rangle^{(0)}$ and the self-consistent estimator $\langle\hat{Q}\rangle^{(s)}$ for both A-type and TM blinking.

\subsubsection*{A-type blinking [Fig. 4(a)]}

The simulation setup for A-type blinking in Fig. 4(a) is identical to that described in S2.3, with the following modifications. The Auger rate is fixed at $k_\textrm{A} = 20 \, k_\textrm{r}$. The OFF-state fraction $p_\textrm{OFF}$ is varied from $0.1$ to $0.9$ in steps of $0.2$ (five values). For each value of $p_\textrm{OFF}$, $10$ independent simulations were performed, each run for $10^4$ switching events. Only simulations whose quantum yield deviated from the theoretical value by less than $2\%$ were retained. All other parameters ($k_\textrm{r} = 1/30 \textrm{ns}^{-1}$, $k_\textrm{nr} = k_\textrm{nr}' = 0$, $T_\textrm{switch} = 0.03$ s, bin size $0.01$ s, $N_\textrm{max} = 120$ counts per bin, $N_\textrm{fit} = 1000$, $M_\textrm{max} = M_\textrm{tot}/10$) are as given in S2.3.

\subsubsection*{TM blinking [Fig. 4(b)]}

The simulation setup for TM blinking in Fig. 4(b) uses the model parameters of QD 3.2 from Ref. \citenum{Podshivaylov2023}, as detailed in S2.4. The pre-exponential factor $k_0$ is varied over five values drawn from a log-uniform distribution in the range $10^{7.2} \leq k_0/k_\textrm{r} \leq 10^{8.8}$. The exponent $\alpha = 10$ and all TLS parameters ($s_0$, $s_i$, $p_i$, $\Gamma_i$) are held fixed at their fitted values from Table S1. For each value of $k_0$, $10$ independent simulations were performed, each run for $5 \times 10^4$ switching events. Only simulations whose quantum yield deviated from the theoretical value by less than $2\%$ were retained. All other parameters (bin size $10$ ms, $N_\textrm{fit} = 1000$, $M_\textrm{max} = M_\textrm{tot}/10$) are as given in S2.4.

\subsection*{S3.2. Parameters for Figure 5: two-dimensional criterion}

Figure 5 of the main text presents the two-dimensional criterion $(\hat{\xi}, \langle\hat{Q}\rangle^{(s)})$ evaluated for A-type, TM, and HC blinking.

\subsubsection*{A-type blinking (colored circles)}

The simulation setup for A-type blinking in Fig. 5 is identical to that described in S2.3, with the following modifications. The Auger rate is set to five fixed values: $k_\textrm{A} / k_\textrm{r} = 0.1$, $1$, $5$, $10$, and $20$ (see legend of Fig. 5). For each value of $k_\textrm{A}$, the OFF-state fraction $p_\textrm{OFF}$ is varied from $0.1$ to $0.9$ in steps of $0.2$ (five values), yielding a total of $25$ distinct parameter sets. For each set, $10$ independent simulations were performed, each run for $10^4$ switching events. All other parameters ($k_\textrm{r} = 1/30 \textrm{ns}^{-1}$, $k_\textrm{nr} = k_\textrm{nr}' = 0$, $T_\textrm{switch} = 0.03$ s, bin size $0.01$ s, $N_\textrm{max} = 120$ counts per bin, $N_\textrm{fit} = 1000$, $M_\textrm{max} = M_\textrm{tot}/10$) are as given in S2.3.

\subsubsection*{TM blinking (triangles)}

The TM blinking data in Fig. 5 were obtained using the multilevel TLS model with parameters taken from two quantum dots reported in Ref. \citenum{Podshivaylov2023}: QD 3.2 (upward triangles) and QD 4.4 (downward triangles).

\textbf{QD 3.2 (upward triangles).} The model parameters are as detailed in S2.4 and Table S1. Two series of simulations were performed:
\begin{itemize}
   \item $k_0$ varied: the pre-exponential factor is set to five values uniformly spaced on a logarithmic scale between $10^{7.2}$ and $10^{8.8}$, with $\alpha = 10$ fixed (red triangles).
    \item $\alpha$ varied: the exponent is varied from $8$ to $12$ in steps of $1$, with $k_0/k_\textrm{r} = 6.17 \times 10^7$ fixed at its fitted value (orange triangles).
\end{itemize}
All other parameters ($s_0$, $s_i$, $p_i$, $\Gamma_i$, $k_\textrm{r}$, $N_\textrm{med}$, bin size, $N_\textrm{fit}$, $M_\textrm{max}$) are as given in S2.4. For each parameter set, $10$ independent simulations were performed, each run for $5 \times 10^4$ switching events.

\textbf{QD 4.4 (downward triangles).} This quantum dot has a lower emission intensity and a non-zero background noise level. The radiative recombination rate is again $k_\textrm{r} = 1/30 \textrm{ns}^{-1}$. The nonradiative rate is $k_\textrm{nr}(t) = k_0 S^{\alpha}(t)$, with the fitted ratio $k_0 / k_\textrm{r} = 7.91 \times 10^5$ and $\alpha = 10$. The Huang--Rhys parameter is modulated by ten independent two-level systems, $S(t) = s_0 + \sum_{i=1}^{10} \sigma_i(t) s_i$, with the parameters listed in Table S2. The mean intensity is $N_\textrm{med} = 40.6$ counts per bin, and the background noise level is $0.53$ counts per bin. Two series of simulations were performed:
\begin{itemize}
    \item $k_0$ varied: the pre-exponential factor is set to five values uniformly spaced on a logarithmic scale between $10^{4.2}$ and $10^{6.8}$, with $\alpha = 10$ fixed (green triangles).
    \item $\alpha$ varied: $\alpha$ from $8$ to $12$ in steps of $1$, with $k_0/k_\textrm{r} = 7.91 \times 10^5$ fixed (cyan triangles).
\end{itemize}
For each parameter set, $10$ independent simulations were performed, each run for $5 \times 10^4$ switching events. All other simulation parameters (bin size $10$ ms, $N_\textrm{fit} = 1000$, $M_\textrm{max} = M_\textrm{tot}/10$) are identical to those used for QD 3.2.

\begin{table}[h]
\centering
\begin{tabular}{c|c|c|c}
\hline
TLS index $i$ & $s_i$ & $p_i$ & $\Gamma_i$ ($\textrm{s}^{-1}$) \\
\hline
-- & $s_0 = 0.0726$ & -- & -- \\
1 & $0.0614$ & $0.318$ & $5.62 \times 10^{-3}$ \\
2 & $0.137$ & $0.239$ & $3.00 \times 10^{-2}$ \\
3 & $0.0244$ & $0.458$ & $1.61 \times 10^{-1}$ \\
4 & $0.274$ & $0.436$ & $8.58 \times 10^{-1}$ \\
5 & $0.0473$ & $0.860$ & $4.58$ \\
6 & $0.0369$ & $0.428$ & $2.45 \times 10^{1}$ \\
7 & $0.0278$ & $0.485$ & $1.31 \times 10^{2}$ \\
8 & $0.0316$ & $0.704$ & $6.99 \times 10^{2}$ \\
9 & $0.0163$ & $0.371$ & $3.73 \times 10^{3}$ \\
10 & $0.0682$ & $0.950$ & $2.00 \times 10^{4}$ \\
\hline
\end{tabular}
\caption{Parameters of the two-level systems modulating the Huang--Rhys parameter for QD 4.4 from Ref. \citenum{Podshivaylov2023}.}
\label{tab:S2}
\end{table}

\subsubsection*{HC blinking (green squares)}

The simulation setup for HC blinking in Fig. 5 is identical to that for A-type blinking described in S2.3, with the following modifications. The decay rate is kept constant at $\Gamma = 1/30 \textrm{ns}^{-1}$ (the same as $\Gamma_\textrm{ON}$ for A-type). Only the intensity switches between $N_\textrm{ON} = 120$ and $N_\textrm{OFF} = 12$ counts per bin. The OFF-state fraction $p_\textrm{OFF}$ is varied from $0.1$ to $0.9$ in steps of $0.2$. All other parameters ($T_\textrm{switch} = 0.03$ s, bin size $0.01$ s, $N_\textrm{fit} = 1000$, $M_\textrm{max} = M_\textrm{tot}/10$, $10^4$ switching events, $10$ simulations per parameter set) are as given in S2.3.

\bibliography{ref}
\bibliographystyle{rsc}